\documentclass[pdflatex,sn-aps,iicol]{sn-jnl}

\usepackage{graphicx}%
\usepackage{multirow}%
\usepackage{amsmath,amssymb,amsfonts}%
\usepackage{amsthm}%
\usepackage{mathrsfs}%
\usepackage[title]{appendix}%
\usepackage{textcomp}%
\usepackage{manyfoot}%
\usepackage{booktabs}%
\usepackage{algorithm}%
\usepackage{algorithmicx}%
\usepackage{algpseudocode}%
\usepackage{listings}%

\usepackage[usestackEOL]{stackengine}
\usepackage{scalerel}
\usepackage{mwe}
\usepackage{subfig}
\usepackage{tikz}
\usepackage{automated_macros}
\usepackage{multirow}
\usepackage{xcolor,colortbl}
\usepackage{caption}
\usepackage{siunitx}
\usepackage[export]{adjustbox}
\theoremstyle{thmstyleone}%
\theoremstyle{thmstyletwo}%

\theoremstyle{thmstylethree}%

\begin{document}

\title[Data-Driven Surrogate Modeling Techniques to Predict the Effective Contact Area of Rough Surface Contact Problems]{Data-Driven Surrogate Modeling Techniques to Predict the Effective Contact Area of Rough Surface Contact Problems}


\author*[1]{\fnm{Tarik} \sur{Sahin}}\email{tarik.sahin@unibw.de}

\author[2]{\fnm{Jacopo} \sur{Bonari}}\email{jacopo.bonari@dlr.de}

\author[1]{\fnm{Sebastian} \sur{Brandstaeter}}\email{sebastian.brandstaeter@unibw.de}

\author[1,2]{\fnm{Alexander} \sur{Popp}}\email{alexander.popp@unibw.de}

\affil[1]{\orgdiv{Institute for Mathematics and Computer-Based Simulation (IMCS)}, \orgname{University of the Bundeswehr Munich}, \orgaddress{\street{Werner-Heisenberg-Weg 39}, \city{Neubiberg}, \postcode{D-85577}, \state{Bavaria}, \country{Germany}}}
\affil[2]{\orgdiv{Institute for the Protection of Terrestrial Infrastructures}, \orgname{German Aerospace Center (DLR)}, \orgaddress{\street{Rathausallee 12}, \city{Sankt Augustin}, \postcode{D-53757}, \state{North Rhine-Westphalia}, \country{Germany}}}


\abstract{
    The effective contact area in rough surface contact plays a critical role in multi-physics phenomena such as wear, sealing, and thermal or electrical conduction.
Although accurate numerical methods, like the Boundary Element Method (BEM), are available to compute this quantity, their high computational cost limits their applicability in multi-query contexts, such as uncertainty quantification, parameter identification, and multi-scale algorithms, where many repeated evaluations are required.
This study proposes a surrogate modeling framework for predicting the effective contact area using fast-to-evaluate data-driven techniques.
Various machine learning algorithms are trained on a precomputed dataset, where the inputs are the imposed load and statistical roughness parameters, and the output is the corresponding effective contact area.
All models undergo hyperparameter optimization to enable fair comparisons in terms of predictive accuracy and computational efficiency, evaluated using established quantitative metrics.
Among the models, the Kernel Ridge Regressor demonstrates the best trade-off between accuracy and efficiency, achieving high predictive accuracy, low prediction time, and minimal training overhead-making it a strong candidate for general-purpose surrogate modeling.
The Gaussian Process Regressor provides an attractive alternative when uncertainty quantification is required, although it incurs additional computational cost due to variance estimation.
The generalization capability of the Kernel Ridge model is validated on an unseen simulation scenario, confirming its ability to transfer to new configurations.
Database generation constitutes the dominant cost in the surrogate modeling process.
Nevertheless, the approach proves practical and efficient for multi-query tasks, even when accounting for this initial expense.
In summary, this work advances rough surface contact modeling into previously prohibitive application domains by integrating machine learning with efficient numerical simulation.
}




\maketitle

\clearpage

\section{Introduction}\label{sec:intro}
The vast majority of real surfaces, either natural or artificial, only collide within a small set of touching points or 
touching zones when brought into contact under moderate pressure. This defines a so-called effective contact area much 
smaller than the nominal contact area observed at a scale comparable to the surface extension in space. 
The reason for this behavior is that surfaces are 
affected by finely spaced irregularities and deviations from the ideal shape that include scratches, pits, grooves and, in 
general, jaggedness resulting from external factors or machining processes, with characteristic dimensions spanning 
several different scales. These imperfections are referred to, in the literature, as \emph{surface roughness}, for which 
a comprehensive theory relevant to surface metrology can be found in Whitehouse~\cite{whitehouse:1994}.

The correct assessment of the effective contact area is a problem of paramount importance in numerous engineering and physics 
applications, since this quantity governs many coupled multi-physics phenomena happening at the contact interface, either 
mechanical, thermal or electrical. The excellent review presented in the work of Vakis et al.~\cite{vakis:2018} 
accurately describes this complex scenario, referred to by the authors as a true \emph{“paradise of multiphysics”}.

As a direct consequence of the utmost importance of this topic, a vast ecosystem of solution methods has developed over 
the last decades, which could be broadly categorized into analytical and numerical approaches. The first modern 
quantitative approach belongs to the first class and dates back to the seminal work of 
Greenwood and Williamson~\cite{greenwood1966}. They developed the first multi-asperity model for the analysis of the 
relationship between real contact area and normal reaction force in a rough contact, and paved the way for all the 
subsequent analytical models whose detailed descriptions would, unfortunately, fall outside the scope of the presented 
publication. The interested reader is referred to Barber~\cite[\S 6]{barber:2000} for a thorough list and detailed 
breakdown of the most important analytical models that have been developed during the last eighty years.
These models have all proven to be excellent tools to gain a deep and fundamental understanding of the problem under examination, 
and they are still used nowadays given their accuracy and speed of deployment. Nonetheless, they are characterized by 
some intrinsic limitations that make them unsuitable concerning various specific contexts. Analytical models are limited 
to derive results in a global sense, meaning that they can not deliver local information for specific points of the 
interface and, furthermore, have to rely on strong assumptions regarding surface geometry. To overcome their limitations, and 
thanks to the dramatic increase in terms of computational resources available, several numerical models have been 
developed parallel to analytical models for approximating solutions of rough contact problems.
As of today, three main approaches, two based on continuum mechanics and one on atomistic models, are commonly used as 
computational workhorses for the numerical treatment of rough contact problems. Within the first category fall the 
Boundary Element Method (BEM), developed by Johnson~\cite{johnson:1985} and specifically elaborated for rough contact 
by Xu and Jackson~\cite{xu:2019}, and the Finite Element Method (FEM), for which related comprehensive computational 
contact schemes have been proposed by Wriggers~\cite{wriggers:2002}, Popp et al. \cite{Popp2010a}, Wriggers and Popp~\cite{wriggers:2018} for example
with suitable numerical algorithms and solvers contributed by Adams~\cite{Adams2004a},
Wiesner et al.~\cite{Wiesner2018a,Wiesner2021a} or Mayr and Popp~\cite{Mayr2023a}.

In BEM models, the rough contact problem is cast as the equivalence between the interface displacement field and a 
Boundary Integral Equation (BIE) coincident with the spatial convolution over the contact domain between a Green's 
function and an unknown contact traction field.
In BEM, the problem complexity is reduced by one dimension, thus making it a very efficient tool, capable of delivering 
very accurate information on local displacements and contact tractions, see for example Pohrt and Li~\cite{pohrt:2014}, 
Vollebregt~\cite{vollebregt:2014}, or Bemporad and Paggi~\cite{bemporad:2015} for a review of different possible 
implementations of the method. Since global quantities can be directly derived, it also proved itself a suitable 
tool for the comparison of the results delivered by different analytical models. Systematic research on this subject 
can be found in Paggi and Ciavarella~\cite{paggi:2010}, Carbone and Bottiglione~\cite{carbone:2008}, and Yastrebov 
et al.~\cite{yastrebov:2017}. Despite this method being capable of analyzing any kind of surface without a priori
assumptions on the asperities distribution, one drawback is that it is mostly limited to infinitesimal
strain and linear material laws. Extensions of the method can be found to account for bulk plasticity, 
viscoelasticity or generic conditions for which a Green's function is not available in analytical form, 
see e.g. Jacq et al.~\cite{jacq:2002}, Putignano and Carbone~\cite{putignano:2014}, and Zhao et al.~\cite{zhao:2016} 
for respective exemplary instances of these three cases. They unfortunately remain limited niche exceptions.

On the other hand, the FEM is much more versatile, since it can deal with the desired material law, 
finite deformations, finite size domains and complex boundary conditions. As a drawback, every part of the domain has 
to be discretized, so that the higher computational cost must be justified by specific requirements for the problem 
to be solved. The last twenty years have seen a continuous growth of the use of the FEM applied to the context under 
examination. In this regard, the first example can be traced back to Hyun et al.~\cite{hyun2004} for the adhesiveless 
frictionless contact between elastic bodies with rough surfaces, later extended by Pei et al.~\cite{pei:2005} to 
account for an elastoplastic response. The FEM has also been employed by Yastrebov 
et al.~\cite{yastrebov:2011} for the solution of rough contact with finite plastic deformations and in Carvalho 
et al.~\cite{carvalho:2023} for the solution of Signorini problems involving deformable bodies in contact with 
geometrically complex rigid surfaces. Recently, a new method has been developed by Paggi and Reinoso~\cite{paggi:2020} 
for a simplified yet accurate analysis of rough contact problems using FEM, in which the rough surface needs not to be 
explicitly modeled, but its shape is embedded in a layer of interface finite elements. The method, named  
\emph{MPJR} (eMbedded Profile for Joint Roughness), was originally designed for the solution of two-dimensional 
adhesive rough contact problems, and has later been extended in Bonari et al.~\cite{bonari:2021} to include friction 
and in Bonari et al.~\cite{bonari:2022} to solve three-dimensional contact. The framework aims at bypassing some 
shortcomings that are faced when solving rough contact problems using FEM, namely the care required in the mesh 
generation process to avoid negative Jacobians and the difficulty related to the explicit discretization of the
rough geometry. 

Molecular Dynamics (MD) methods fall within the last category of numerical models. They can trace the kinematics of 
every single atom by computing its interaction with all particles in the domain, using a predefined 
potential. Since every atom has to be explicitly modeled, the dynamic interaction time scales are inherently low, 
and the method is therefore particularly suitable for the analysis of very limited space and time domains; 
cf.~\cite[\S2.5]{vakis:2018}. To overcome this, the elastic potentials of a single layer of atoms can be normalized to 
incorporate the elastic interactions proper to a semi-infinite solid. This Green's Function Molecular Dynamics (GFMD) approach, 
developed by Campa\~{n}{\'a} et al.~\cite{campana:2006}, proved to deliver very accurate solutions, as demonstrated by 
the results of the recent contact mechanics challenge issued by M\"{u}ser et al.~\cite{mueser:2017}.

To leverage the advantages of one method over its drawbacks, multi-scale approaches can be defined, in which different 
models are employed for the solution of different scales of the problem, with information exchange enforced across the 
scales. In Bonari et al.~\cite{bonari:2020b}, FEM is used at the macro-scale, where a perfectly smooth interface is 
considered, while BEM is employed for the solution of a micro-scale rough contact problem. A bidirectional coupling 
between the two scales is enforced in terms of mean-plane separation.

Despite the huge steps forward in numerical solution schemes, and even though the available computational resources 
continuously increased, the high-fidelity numerical solution of a rough contact problem can still be cumbersome. 
An excellent example are parameter studies, where numerous simulations have to be run, causing the 
computational cost to grow rapidly and eventually become prohibitive. 
For instance, in the context of optimizing material properties or surface topographies for improved contact performance, 
multiple simulations are required to evaluate different design parameters, making high-fidelity modeling computationally 
expensive.

In this regard, a possible solution is offered by the huge advancement and development experienced in the last few 
years by efficient surrogate modeling techniques. Among them a promising approach, 
namely data-driven surrogate modeling, is 
based on experimental or synthetically generated data, thus deliberately excluding a physics-based modeling of the system; instead, 
the model extracts underlying patterns in provided data by deploying machine learning algorithms.
Many application cases of data-driven surrogate modeling techniques specifically 
crafted for the realms of science and engineering are available in the literature~\cite{Solomatine2008, Montans2019, Bourdeau2019, Karapiperis2021}.
These techniques aim to achieve high accuracy while requiring minimal computational or prediction time. 

Data-driven surrogate modeling techniques can be developed for various purposes such as prediction, 
parameter identification, classification, and anomaly detection. When focusing on rough surfaces, 
these techniques find application across a diverse range of scenarios in fluid mechanics, tribology, and contact mechanics. 
Sanhueza et al.~\cite{Sanhueza2023} deployed machine learning as a parameter identifier to predict the behavior
of turbulent flows past rough surfaces. They applied convolutional neural networks on the height map of a rough surface 
to identify the corresponding skin friction values and Nusselt numbers. 
Elangovan et al.~\cite{Elangovan2015} employed multiple linear regression to identify surface roughness in turning 
operations by considering machining parameters such as speed, depth of cut and tool wear. The resulting database 
comprises statistical features derived from vibration signals, and, in addition, principal component analysis is 
applied to reduce the dimensionality of the input space. 
In the study by Prajapati et al.~\cite{Prajapati2023} the researchers constructed a surrogate model utilizing a 
regression-based Multi-Layer Perceptron  (MLP) to predict elastohydrodynamic lubrication parameters, specifically the traction coefficient and asperity load ratio. These predictions were customized for different surface topographies 
encountered in rough contact problems. Similarly, Boidi et al.~\cite{Boidi2020} applied the machine learning radial basis 
function method to identify the coefficient of friction on textured and porous surfaces. In addressing rough surface 
problems, there are numerous additional examples where machine learning techniques have been applied to identify parameters 
associated with surface roughness, lubrication and wear. This is well-documented in the reviews by Syam~\cite{Syam2022} 
and Batu~\cite{Batu2023}. However, only a few studies have delved into predicting the mechanical response 
associated with contact problems on rough surfaces, such as the prediction of effective contact area and contact force. 
In the work led by Rapetto et al.~\cite{Rapetto2009}, the authors explored the ability of neural networks to 
predict the relation between roughness parameters and the effective contact area. The generated dataset comprises 
statistical features derived solely from 2D profiles, with the authors noting that the error tends to be higher for 
small values of the effective contact area. 
Similarly, Kalliorinne et al.~\cite{Kalliorinne2021} explored the ability of multitask neural networks to predict 
the mechanical response of contact, including the effective contact area and contact pressure, across surface 
topographies generated using distinct methods based on a given heights probability distribution and power spectrum.
In research conducted by Jiang et al.~\cite{Jiang2021}, a new method was proposed to predict the effective contact area of fractal rough 
surfaces. This method employs a regression model based on least-squares support vector machines.
As the fractal surfaces are generated using a deterministic Weierstrass-Mandelbrot (WM) model,
the resulting database does not include the statistical features but comprises solely the roughness parameters 
inherent to the WM model and the external load.
Furthermore, the generated database has a very limited range due to costly FEM simulations. Additionally,  
they do not report generalization errors on the unseen data since only a single sample is left for testing. 
Another approach for generating a surrogate model, namely Physics-Informed Neural Networks (PINNs), 
incorporates the physical laws of a governing problem and the additional data into the loss function of the 
neural networks~\cite{raissi2019}. For instance, in the context of contact mechanics, Sahin et al.~\cite{sahin2024} deployed 
PINNs as a fast-to-evaluate surrogate model to solve forward and inverse contact mechanics problems on the macro-scale. 

As stated above, only very few studies have focused on predicting the effective contact area 
in contact problems involving rough surfaces, employing state-of-the-art data-driven techniques. 
Therefore, there is a continuous demand for the development of such techniques.  
In this manuscript, we develop a surrogate modeling framework to predict the effective contact area stemming from the 
contact between a linear elastic half-space and a rigid rough surface. Rough topographies are generated utilizing the 
RMD algorithm to obtain self-affine fractal surfaces. The generated database includes 
a wide range of parameters, viz. model parameters, material parameters and roughness parameters obtained through 
statistical characterization of topographies. These statistical roughness parameters are calculated based 
on the height field, summits (maxima in 3D) and peaks (maxima in 2D) of fractal surfaces. In contrast to earlier 
investigations in the literature, various machine learning algorithms are deployed as fast-to-evaluate models utilizing a 
database generated with Boundary Element Method simulations. As mentioned earlier, 
BEM stands out 
as an efficient approach, providing highly accurate results with significantly reduced computational time compared to 
other numerical schemes. 
Thus, we couple state-of-the-art data-driven methodologies with enriched simulation data based on BEM. 
These methodologies include 
Decision Tree, Random Forest, Support Vector Machine (SVM) Regressor, Adaptive Boosting (AdaBoost),
Gaussian Process (GP) Regressor, Gradient Boost, Kernel Ridge, eXtreme Gradient Boosting (XGBoost), 
and Multi-Layer Perceptron (MLP) Regressor.
Hyperparameter optimization is performed to identify the best configuration for each model type.
This optimization is carried out through grid search combined with cross-validation to 
identify the optimal model parameters.
For quantitative comparisons among surrogate models, several accuracy metrics are computed. 
These metrics include the normalized mean square error (nMSE), normalized mean absolute error (nMAE), 
normalized max-error (nMaxE) and the $R^2$ score.
Selecting an optimal surrogate model requires balancing accuracy and prediction performance, depending on the specific requirements of the application task.  
Our results indicate that kernel-based regressors generally perform very well in the given context of rough surface contact simulations.
Among them, the Kernel Ridge model offers the best trade-off between accuracy and computational efficiency.  
It achieves high predictive accuracy, low prediction time, and minimal training overhead, making it a strong candidate for general-purpose surrogate modeling.  
Therefore, the Kernel Ridge model is eventually compared to the BEM model in terms of overall computational time. 
This assessment encompasses not only the prediction time 
of the trained model but also considers the training, tuning, and data generation time. Going one step further, to assess the model's 
generalization capability, the optimal model is tested in a new simulation scenario, which is not seen during the training 
process. The findings reveal that the developed surrogate model performs well in this 
previously unseen scenario. 
Specific scientific contributions of the paper can be outlined as follows:
\begin{itemize}
    \item Development of an efficient fast-to-evaluate surrogate modeling workflow to predict the effective contact area 
    for rough surface contact problems.
    \item Establishment of an informative database encompassing a diverse set of parameters combined with BEM simulations.
    \item Generalization of the machine learning model by testing on the unseen simulation scenario.
\end{itemize}  

The remainder of this article is structured as follows. Section~\ref{sec_problem_formulation} describes the problem 
formulation of rough surface contact between a linear elastic half-space and a rigid rough surface generated by the 
Random Midpoint Displacement (RMD) algorithm. An efficient numerical approach, the BEM, is outlined as a state-of-the-art method 
to compute the effective contact area. Also, a variety of statistical parameters are used to describe surface features, 
thus creating a meaningful database for surrogate modeling. In Section~\ref{sec_methods}, we elucidate a methodology for 
surrogate modeling as an alternative to conventional techniques. This includes the development of a surrogate modeling workflow, 
the generation of a database exploiting BEM simulations, and the introduction of fast-to-evaluate surrogate models as 
baseline models and tuned models via hyperparameter optimization. Furthermore, several accuracy metrics are introduced 
to validate the accuracy of surrogate models. Section~\ref{sec_results} covers an evaluation of baseline and tuned models 
comparing them in terms of accuracy and performance. Among the investigated machine learning approaches, the so-called optimal model is 
compared with the BEM model in terms of performance. Also, the optimal model (Kernel Ridge) is evaluated in a different environment with unseen data. 
Section~\ref{sec_conclusion} concludes the manuscript by summarizing the main findings and providing insights into future 
research directions.

\section{Problem Formulation}\label{sec_problem_formulation}
\subsection{Physical Problem: Rough Surface Contact}\label{sec_rsc}
A rough surface contact problem typically refers to a class of engineering challenges that 
involves the interaction between two surfaces which are conformal at the macro-scale, but non-conformal at smaller scales, given the presence of micro-scale textures, defects, or, in general, asperities~\cite{johnson1985}.

The study of these problems stemmed from the attempt to explain Coulomb's law of friction - 
specifically, the observation that the tangential force $Q$ required for the relative sliding 
of two bodies in contact is proportional to the normal force $P$ that pushes them together, 
and independent of the observable contact area~\cite{bowden1950}. 
However, at the microscopic level, real surfaces are rough and contact occurs only at discrete asperities, forming the effective contact area $\EffectArea$ which is typically much smaller than the nominal contact area. This effective 
contact area plays a crucial role in friction, wear, and sealing mechanisms, as it governs the actual load-bearing 
capacity and energy dissipation at the interface~\cite{vakis2018}.  
Therefore, its correct assessment is essential for the analysis of physical phenomena related 
to friction, wear, and sealing, i.e., the class of problems investigated in the field of tribology~\cite{vakis2018}.  
On top of that, thanks to electro- and thermo-mechanical analogies~\cite{barber2018}, the effective contact area also plays a crucial role in the study of electrical and thermal exchanges at the interface level.

Many analytical models have been developed for the identification of the effective contact area 
in this context, but they rely on strong assumptions for the statistical distribution of the 
elevation field describing the rough surface and, most importantly, they deliver a global 
response that can not describe the solution at every point of the contact domain. In contrast, numerical models can not only estimate the effective contact area but also determine the contact status of each interface point--that is, identify which points are in contact and which are not.

\begin{figure}[b]
    \centering
    \includegraphics[width=\columnwidth]{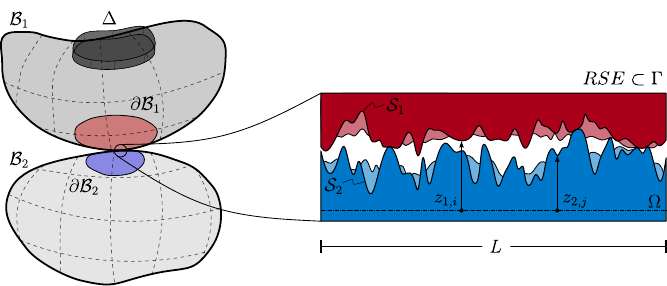}
    \caption{Graphical representation of the problem under consideration. At a macroscopic scale, two deformable bodies come into contact under the action of an external imposed displacement $\Delta$. At the microscopic scale, roughness features can be revealed, resulting in an effective contact area $\EffectArea$ much smaller than the nominal contact area that is assumed at the macroscopic scale.}
    \label{fig:contact_model}
\end{figure}

A common numerical analysis approach is based on the hypothesis of scale separation. Consider the two bodies $ B_i $, $ i \in \{1, 2\} $, represented in Figure~\ref{fig:contact_model}, brought into contact by the action of an external load, which can be an applied force or an imposed act of motion $\Delta$, the latter going under the name of \emph{far-field displacement}. We assume that the shapes of the contacting surfaces $ \mathcal S_i $, which correspond to portions of the boundaries $ \partial B_i $ of the bodies, match or conform to each other, enabling the creation of a finite nominal contact area. In this case, a nominally smooth contact interface $ \Gamma $ can be defined, encompassing $ \mathcal S_i $ along with the underlying portions of the material bulk.

We now assume that the contacting surfaces are microscopically rough, and focus the attention on a subset of $ \Gamma $ small enough to make the roughness characteristics relevant. This subset is referred to as a Representative Surface Element (RSE). In most cases analyzed in the literature, the RSE is modelled as a square patch with size $ L $, discretized on a Cartesian grid with $ n $ points per side and grid size $ g_\mathrm{l} = L/(n-1) $. On this grid, roughness is described by two discrete height fields $ \boldsymbol{z}_i \in \mathbb{R}^{n \times n} $, i.e., a height value for both rough surfaces is assigned to every node of the grid, calculated from a reference datum plane $ \Omega $ for each of the surfaces $ \mathcal S_i $.

If the macroscopic, nominally smooth contact interface $ \Gamma $ is flat, $ \Omega $ can be considered as part of $ \mathcal S_i $. Otherwise, it can be regarded as the tangent plane at the point under consideration: given the hypothesis of scale separation, we assume that the local radius of curvature $ R $ satisfies $ R >> L $, where $ R $ is the common local radius of curvature for both $ S_i $.

\begin{figure}[tbhp] 
    \centering
    \begin{tikzpicture}
        \node[inner sep=0pt] (problem) at (8.5,0.5)
            {\includegraphics[trim={0cm 0cm 0cm 8cm}, clip,width=1.1\columnwidth]{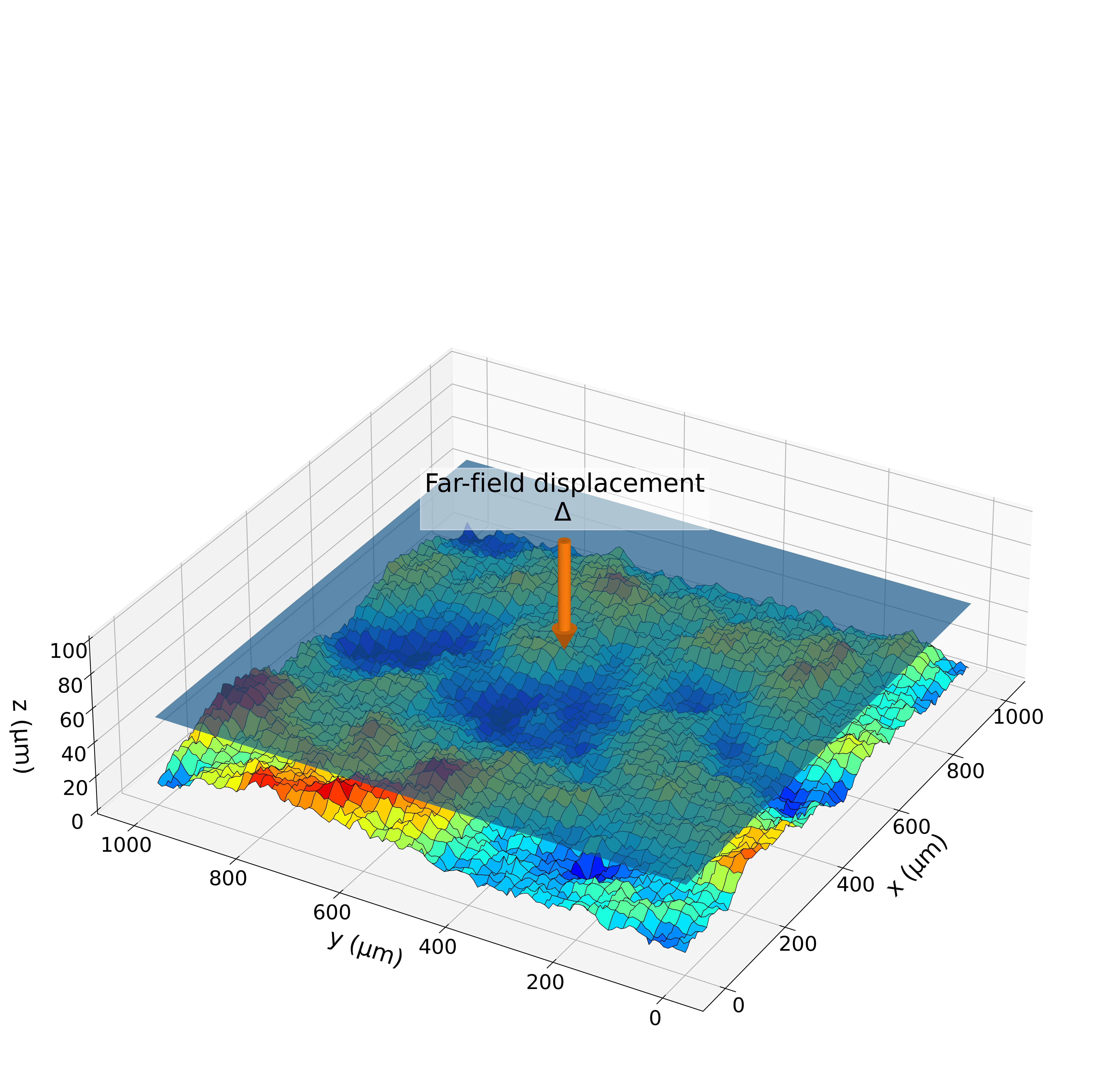}};
            \node[inner sep=1pt] (elastic) at (11,2.5)
            {\footnotesize Elastic half-space};
        \draw[->,thick] (10,1) -- (elastic.south)
            node[midway] {};
        \node[inner sep=1pt] (rigid) at (13,1.5)
            {\footnotesize Rigid rough surface};
        \draw[->,thick] (11,0) -- (rigid.south)
            node[midway] {};
        \end{tikzpicture}
\caption{ 
    Schematic depicting the standard rough surface contact problem.
    Since a composite formulation is applied, the rough surface is assumed to be rigid while the flat surface is linear elastic with composite material properties.}
\label{fig:rsc_problem}
\end{figure}
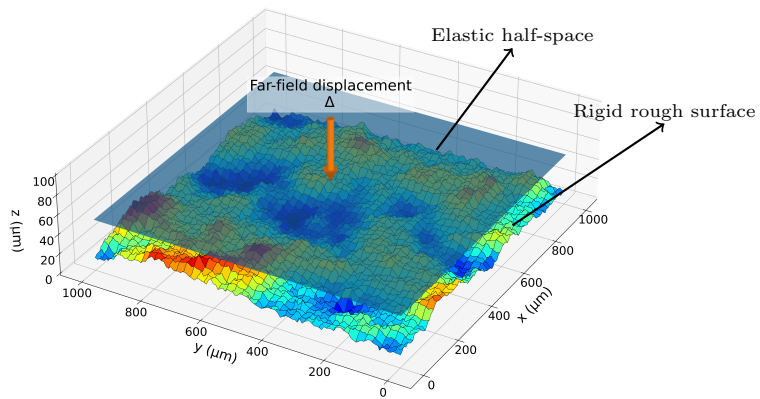

If this approach is chosen, a variable for the number of points in contact, $n_c$, can be introduced and the effective contact area for the RSE expressed as:
\begin{equation}
    \EffectArea = 100 \frac{n_c \ g_l^2}{L^2}\%,
\end{equation}
If the bodies in contact are linear elastic, the solution of the contact problem in terms of $\EffectArea$, or, equivalently, $n_\mathrm c$, is only governed by the relative normal displacements $\boldsymbol u_i$ of opposite points belonging to the two sides of $\Gamma$~\cite[\S 2.2]{barber2018}.
Thanks to this property, the problem can be simplified and re-formulated as a contact problem involving a rough rigid surface characterized by a \emph{combined height field} $\boldsymbol z = \boldsymbol z(\boldsymbol z_i) \in \mathbb R^{n\times n}$ and a semi-indefinite linear elastic half-space with \emph{composite mechanical parameters} $E = E(E_i)$ and $\nu = \nu(\nu_i)$ cf. Figure~\ref{fig:rsc_problem}. However, given the assumptions of linearity, and since for small values of $\EffectArea$ a well acknowledged result in tribology is the proportionality between contact area and imposed load, it can be concluded that the actual pattern of contact spots is dictated by the surfaces topography only, and not by the value of the composite Young's modulus.

\begin{figure}[t]
    \centering
    \subfloat[Reference model.\label{subfig:physics}]
    {\includegraphics[valign=c,width=0.3\textwidth]{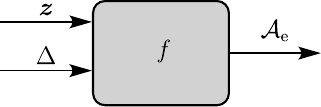}}
    \hspace{0.1\textwidth}
    \subfloat[][Surrogate model.\label{subfig:surrogate}]
    {\includegraphics[valign=c,width=0.3\textwidth]{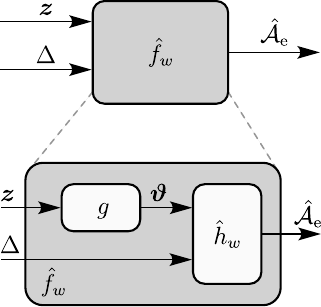}}
    \caption{Surrogate modeling framework for the effective contact area in rough surface contact:
    \protect\subref{subfig:physics} the reference model $f$ computes $\mathcal{A}_\mathrm{e}$ as a function of far-field displacement $\Delta$ and surface height field $\HeightField$, requiring costly numerical simulations;
    \protect\subref{subfig:surrogate} the surrogate model $\SurrogateModel$ approximates this using a fast-to-evaluate regression model $\hat{h}_{\SurrogateParameters}(\Delta, \Statistics)$, where $\Statistics = g(\HeightField)$ is a vector of statistical parameters. The function $g$ is a known deterministic mapping that enables dimensionality reduction of the input vector.}
    \label{subfig:jac}
\end{figure}

In summary, the key parameters governing the computation of the effective contact area $\EffectArea$ are the imposed load $\Delta$ and the surface height field $\boldsymbol{z}$.  
Thus, in abstract terms, solving the problem means evaluating the following function:

\begin{equation}
\label{eq:true_model}
    \EffectArea = f(\Delta, \boldsymbol z).
\end{equation}
This problem setup is illustrated in Figure~\ref{subfig:physics}.  
We know that such a functional relationship exists~\cite{}, and that approximate well-known general closed-form expressions for $f$ are available from many different classical analytical models. However, a lack of overall general agreement in terms of outcome is experienced, with different theories leading to slightly but non-negligible differences in their findings~\cite{YASTREBOV201583}.
Furthermore, the derivation of closed form expressions comes with the cost of strong starting assumptions, e.g., in terms of the topography that must be considered.
Instead, evaluating the function $f$ without a priori simplifications requires the numerical solution of a boundary integral equation that models the whole contact mechanics problem.
As described in Section~\ref{sec:intro}, various numerical schemes can be employed to solve this problem.  
In this work, we use the Boundary Element Method (BEM) for which further details will be provided in Section~\ref{sec_bem}.  

\subsection{Random Midpoint Displacement Algorithm}\label{sec_rmd}
\begin{figure}[b]
    \def\figsize{0.6}
    \def\hskiplocal{\hskip 0.cm}
    \centering
    \begin{tabular}{c@{}c}
    \stackinset{c}{}{t}{-.1in}{ grid size $2\times 2$}{\includegraphics[width=\figsize\columnwidth]{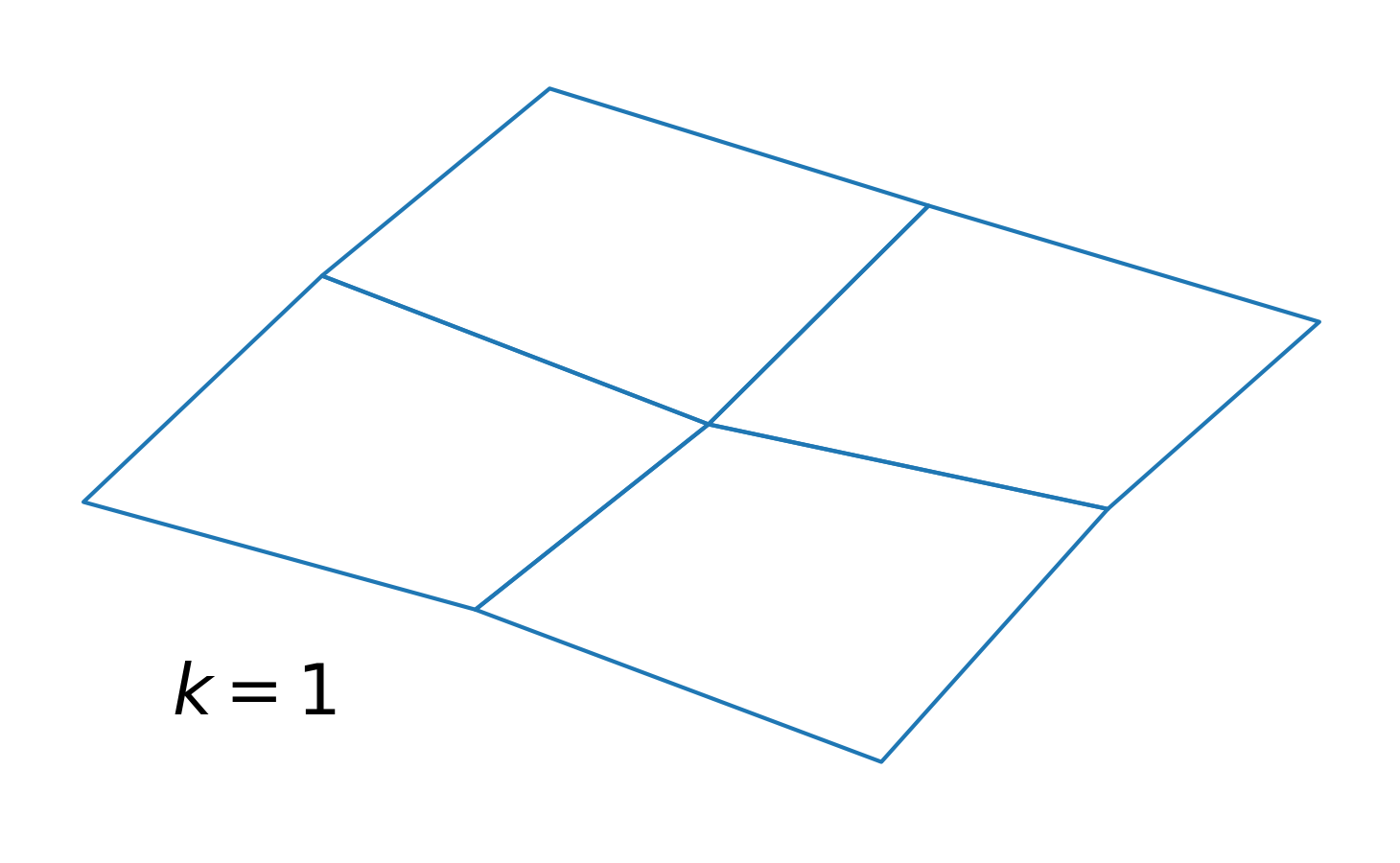}} &
    \stackinset{c}{}{t}{-.1in}{ grid size $8\times 8$}{\includegraphics[width=\figsize\columnwidth]{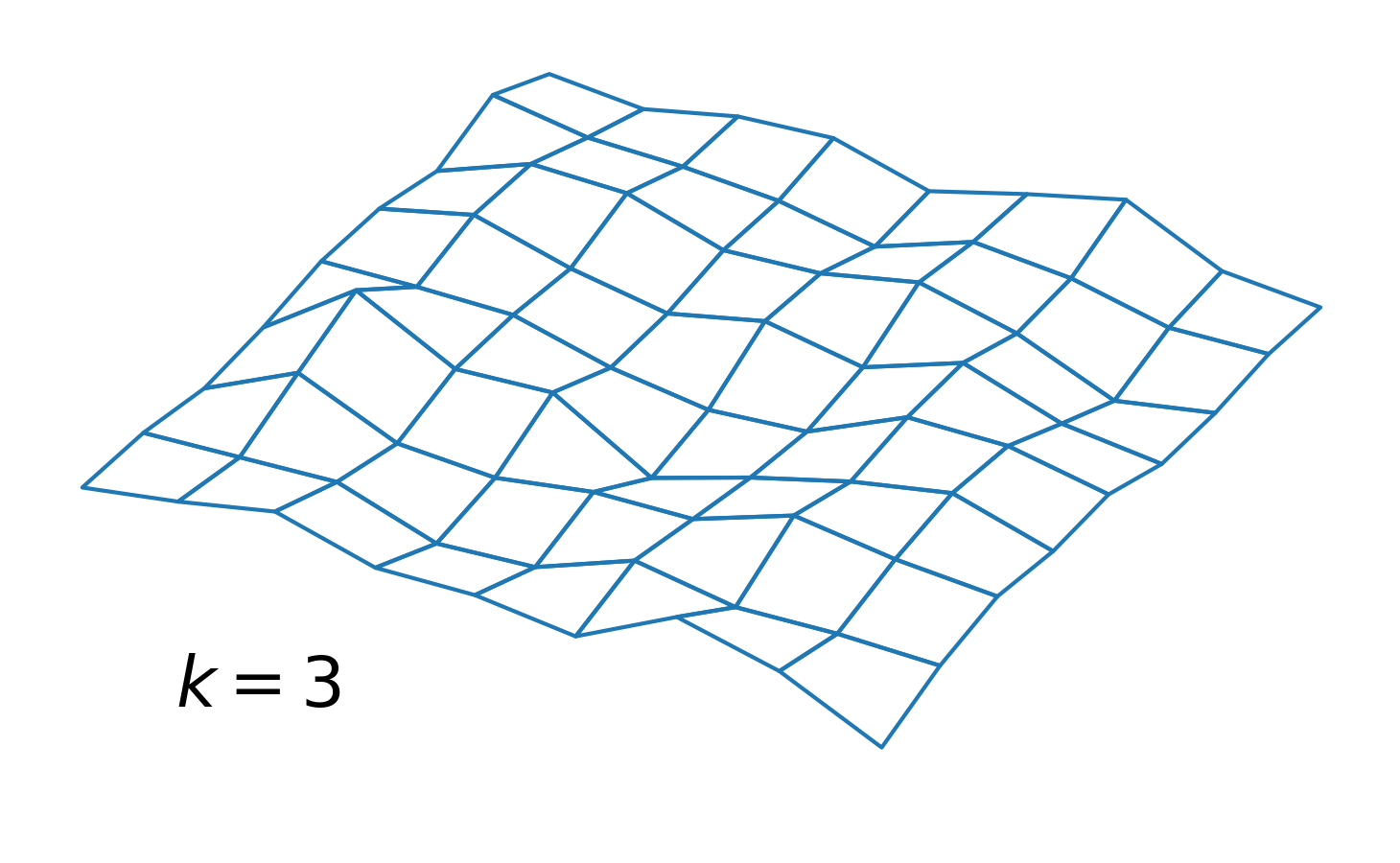}}
    \\[1ex]
    \stackinset{c}{}{t}{-.1in}{ grid size $32\times 32$}{\includegraphics[width=\figsize\columnwidth]{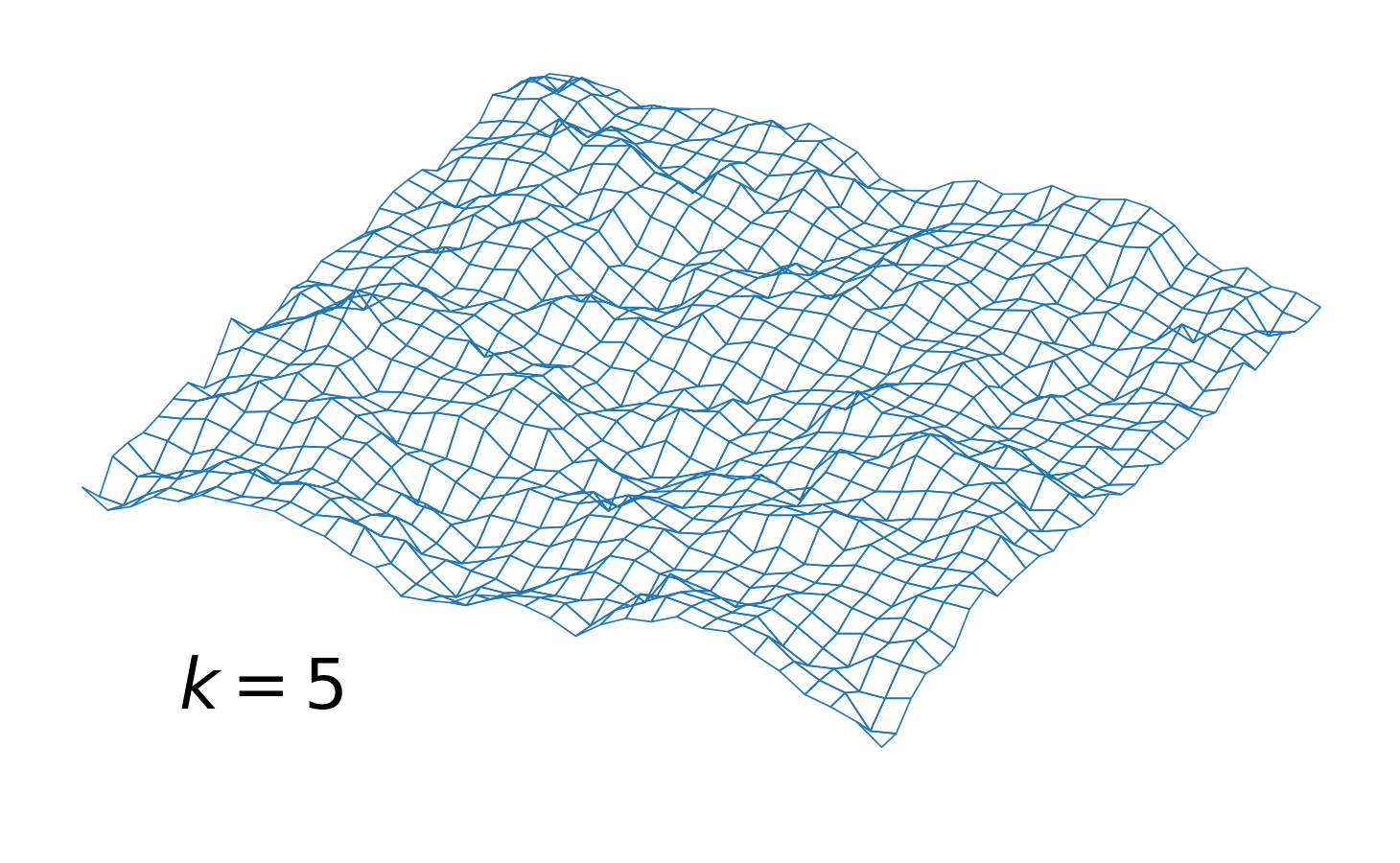}} &
    \stackinset{c}{}{t}{-.1in}{ grid size $128\times 128$}{\includegraphics[width=\figsize\columnwidth]{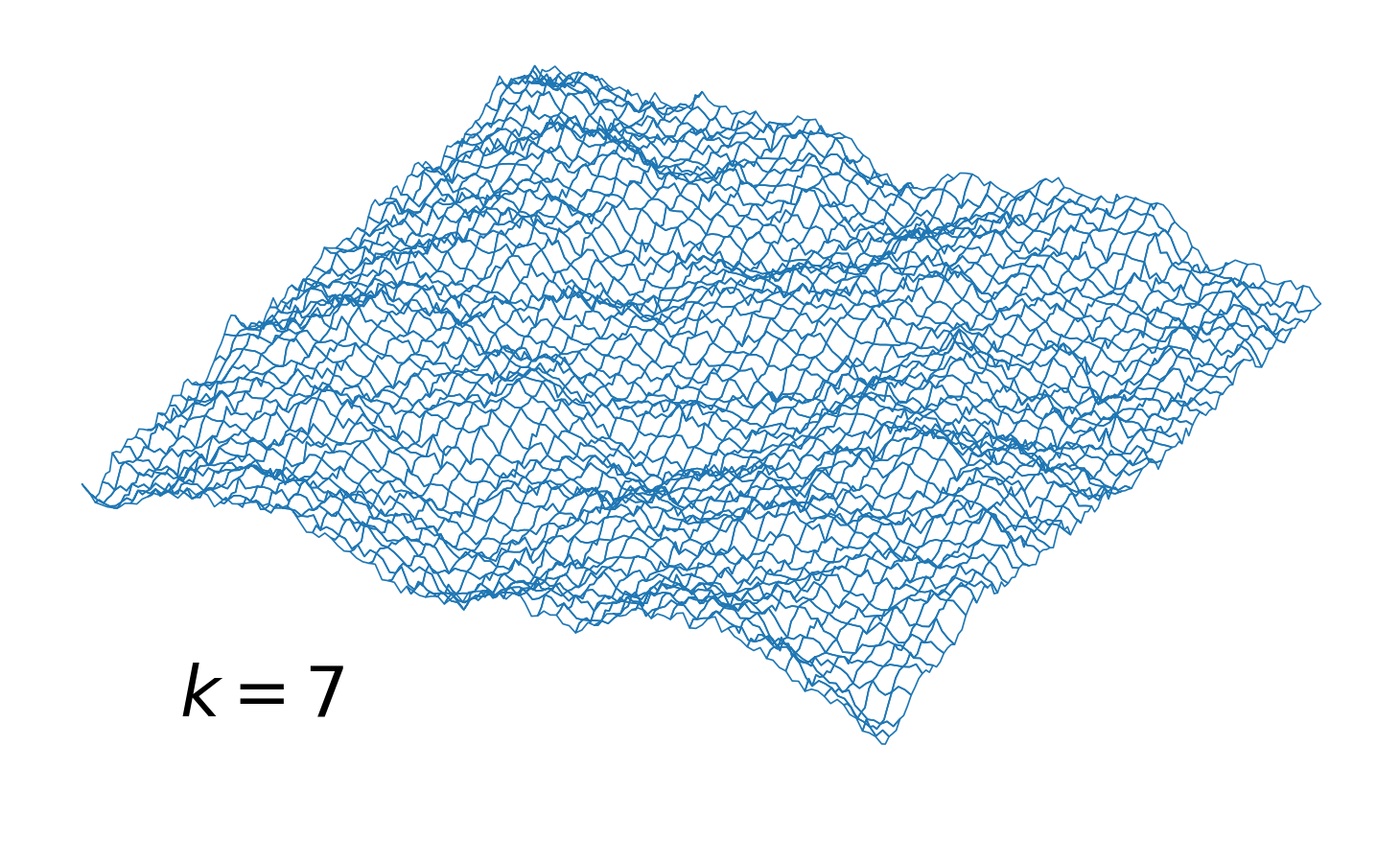}}
    \end{tabular}
    \caption{An illustration of the \textit{Random Midpoint Displacement} (RMD) algorithm. A higher interaction count $k$ results in a finer discretization of the generated self-affine fractal surface.}
    \label{fig:rmd}
\end{figure}

In this study, we are interested in predicting the effective contact area for a given rough surface, encoded by its height profile $\boldsymbol{z}$.  
We use the Random Midpoint Displacement (RMD) algorithm to generate synthetic rough surfaces.  
RMD is a well-established method for producing surfaces with realistic properties.  
Originally introduced by Fournier et al.~\cite{Fournier1982}, it is commonly used to generate self-affine fractal surfaces~\cite{persson2014}.  
This algorithm is a recursive process that starts from the known heights of four points located at 
the corners of a square grid made of a single cell with lateral side $L$. 
The cell is then split in four, resulting in a grid with nine nodes and four smaller cells with side 
$L/2$. Then, a height is assigned 
to each of the five newly created nodes. The height of the four nodes on the edges is given by the 
mean value of the heights of the two adjacent corner nodes plus a random perturbation extracted 
from a Gaussian distribution with zero mean and standard deviation $\sigma_0$. 
The height of the central 
node is given by the mean value of the four corner nodes plus a random perturbation extracted 
from the same Gaussian distribution $\mathcal{N}(0, \sigma_0)$ used for the edge nodes. 
The process is repeated as many times as desired, and for every 
iteration the standard deviation associated with the Gaussian distribution is scaled 
by a factor $2^{-H}$, where $H$ is the generalized Hurst exponent.
In this study, a range between 0.5 and 0.8 is chosen for $H$.  
This range includes typical values observed in many artificial surfaces~\cite{persson2014} and in various natural phenomena that exhibit fractal self-affinity~\cite{mandelbrot1984, barnsley1988, bemporad2015}. 
The key point of the algorithm is the scaling law of the standard deviation: 
it can be proven that for this value the height variation between grid 
points separated by a distance $1/2^n$ is proportional to $1/2^{nH}$, as required by the definition 
of self-affine processes~\cite{barnsley1988}. 
An illustrative example of how the RMD algorithm works is given in Figure~\ref{fig:rmd}. In the current study, surfaces with both $128\times128$ and $256\times256$ elements per side have 
been investigated, thus employing seven and eight iterations of the RMD algorithm, respectively.

\subsection{Statistical Characterization of Topographies}\label{sec_stat_of_top}
Statistical characterization of rough surfaces plays an important role in identifying hidden 
patterns~\cite{zavarise2004}. This result has been achieved by applying the Random Process Theory (RPT), 
previously employed in the study of random noise signals, to the surfaces under examination. 
Functional relationships can be drawn between the statistical parameters of the surfaces and the features of 
the related contact problem. As an example, many analytical models predict that the ratio between 
effective contact area and applied load is proportional to the mean square profile slope, which quantifies the average 
steepness of the surface asperities and is mathematically defined as the expectation of the squared derivative of the 
surface profile~\cite{paggi2010}. 
Furthermore, many legacy asperity-based normal contact models, most of 
them stemming from the pioneering works of Greenwood and Williamson~\cite{greenwood1966}, hinge on the distribution of 
the surface heights.

\begin{figure*}[tbhp]
    \centering
    \begin{tikzpicture}
        \node[inner sep=0pt] (asperity) at (0,0)
            {\stackinset{c}{}{b}{}{(a)}{\includegraphics[trim={0cm 0cm 0cm 8cm}, clip, width=0.45\textwidth]{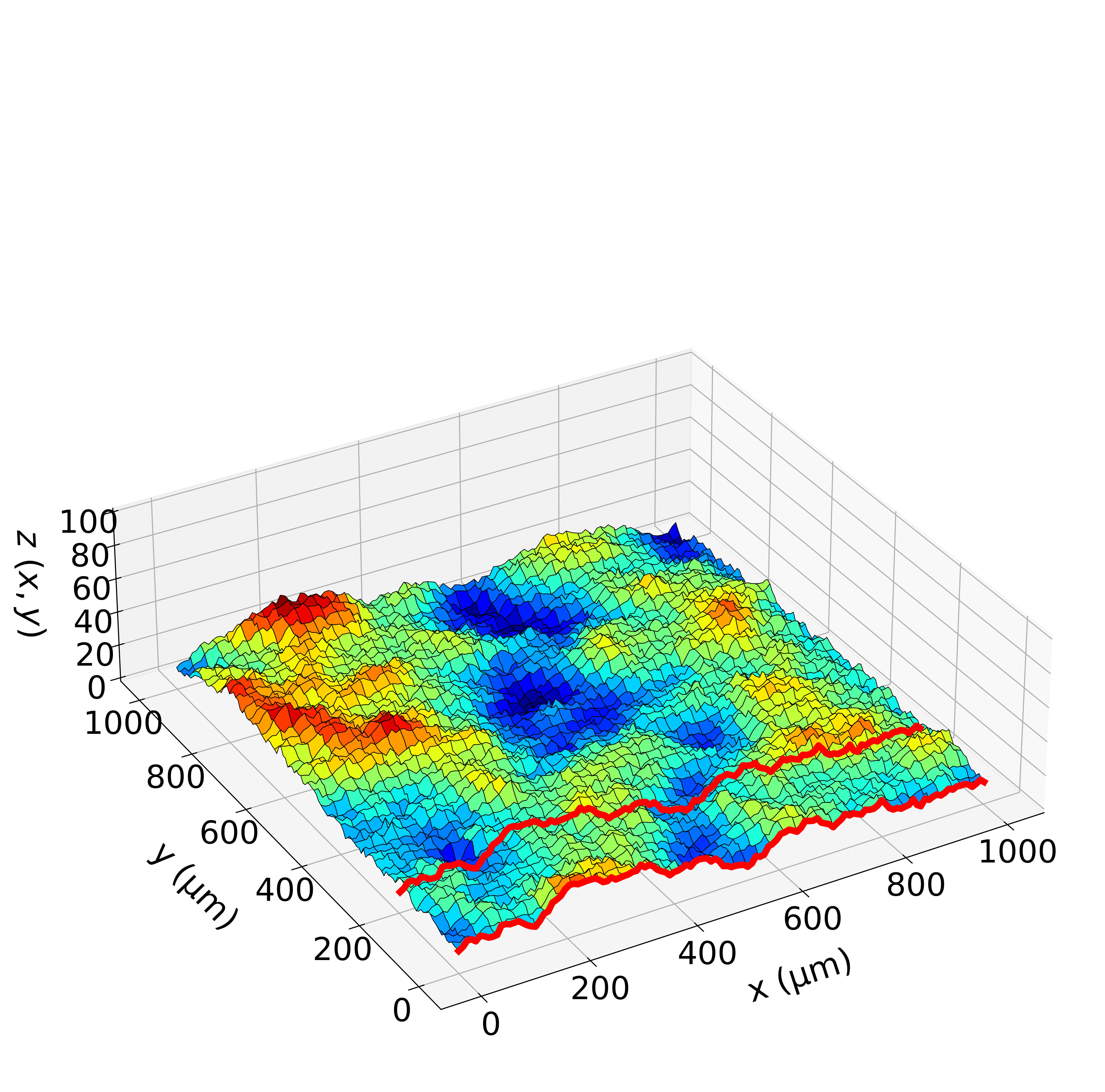}}};
        \node[inner sep=0pt, fill=white, rotate=22.5] (profilometer) at (-0.65,-2.1)
            {\footnotesize Profilometer};
        \draw[->,thick] (-0.75,-2) -- (-1.25,-1.5)
            node[midway] {};
        \node[inner sep=0pt] (profile) at (8.5,0)
            {\stackinset{c}{0.4cm}{b}{-0.5cm}{(b)}{\includegraphics[width=.415\textwidth]{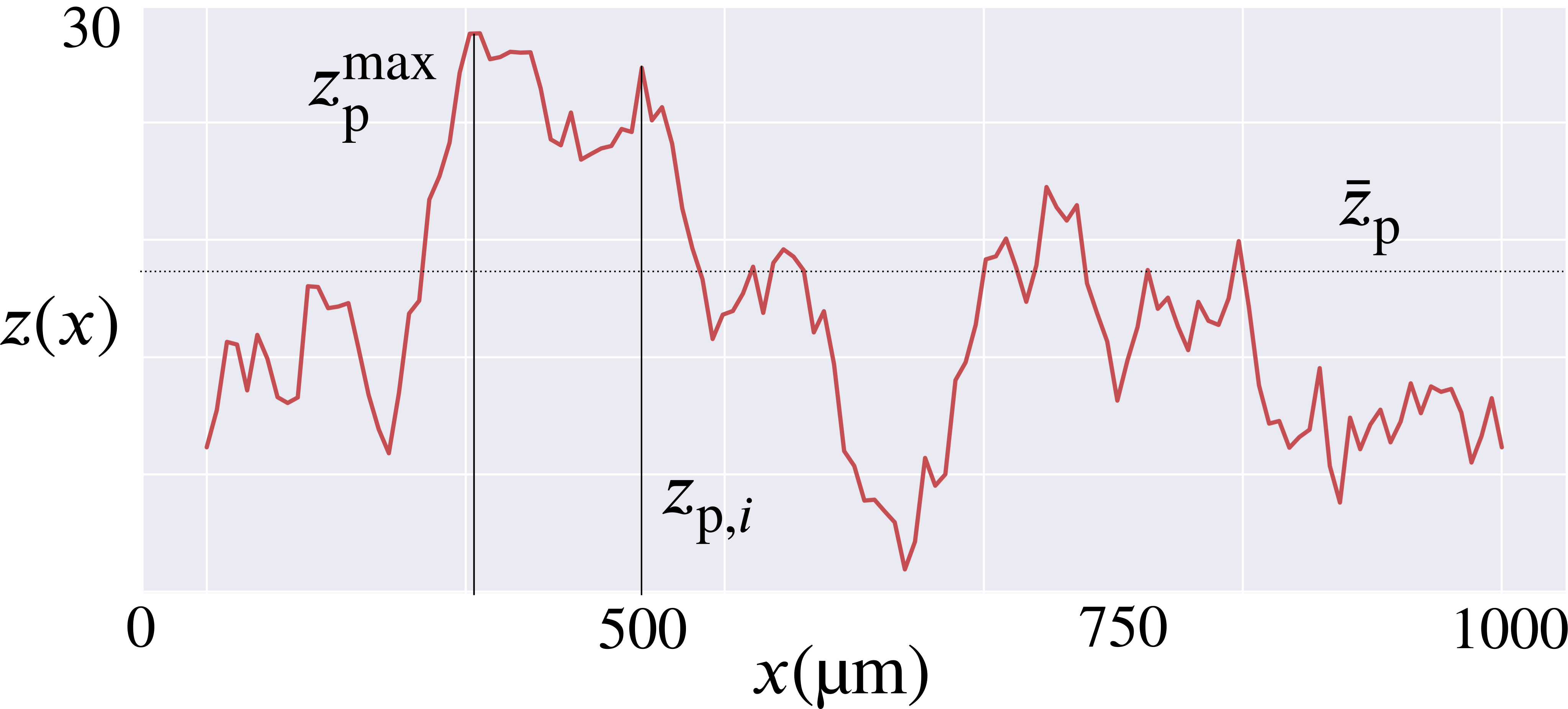}}};
        \draw[->,thick] (2,-0.965) -- (profile.west)
            node[midway] {};
    \end{tikzpicture}
    \caption{(a) 3D representation of a rough topography. (b) A corresponding 2D profile characterized by specific statistical parameters, such as
    peak heights $z_{\mathrm{p},i}$ with accompanying maximum and mean values, $z^\mathrm{max}_\mathrm{p}$, ${\bar{z}}_{\mathrm{p}}$, respectively.}
    \label{fig:stats}
\end{figure*}
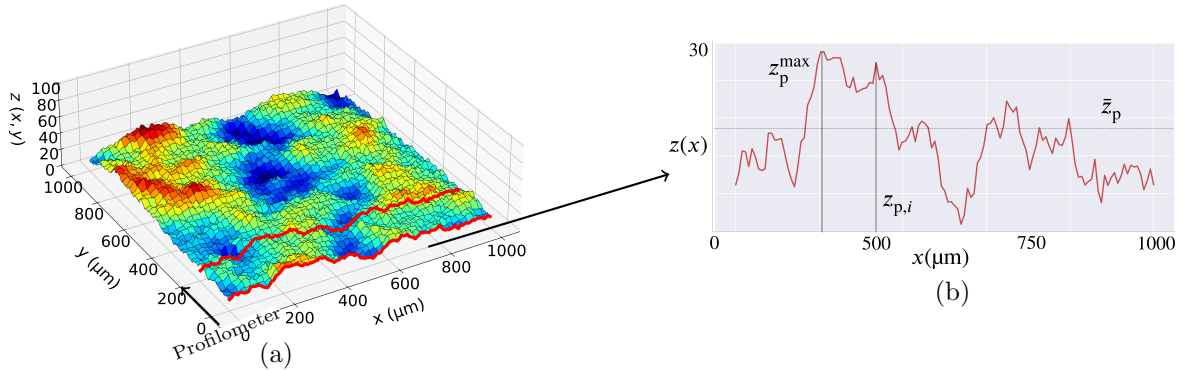

The statistical characterization can be performed by considering the surface as a 3D entity, 
sampling the height field and the summits (maxima in 3D) distributions, or by considering the 
surface as a global assembly of multiple 2D section cuts, or profiles, and then analyzing each 
profile's heights, slopes and peaks (maxima in 2D) distribution~\cite{borri2015}. 
Peaks are measured on profiles generated by a profilometer 
which moves in both the x- and y-direction through the scan length of the topography as depicted in Figure~\ref{fig:stats}a.
A profile can be characterized by peak heights $z_\mathrm{p}$, the maximum of peaks $z^\mathrm{max}_\mathrm{p}$ and
the mean of peaks ${\bar{z}}_{\mathrm{p}}$, see Figure~\ref{fig:stats}b).

In the current work, the statistical parameters usually employed in classical mechanical models will
be derived from the rough surfaces generated according to the procedure presented 
in Section~\ref{sec_rmd} and used as input features for the purely data-driven model as presented 
in Section~\ref{sec_workflow}.
Every quantity derived for this purpose is presented and listed in Table~\ref{tab:list_of_params},
which comprehensively lists all the statistical parameters used in this study.

\subsection{State of the Art: BEM-based Numerical Simulations}\label{sec_bem}
\begin{figure}[b]
    \centering
    \includegraphics[width=\columnwidth]{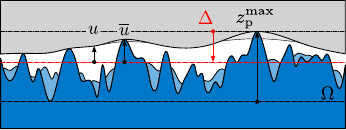}
    \caption{Under the action of the vertical far-field displacement $\Delta$, the deformable bodies are pushed into contact. To properly formulate the LCP problem, an interference $\overline{\boldsymbol{u}}$ can be defined, equal, at each node, to $\boldsymbol{z}-(z_\mathrm{p}^\mathrm{max}-\Delta)$.}
    \label{fig:deformation}
\end{figure}

As stated in Section~\ref{sec_rsc}, the solution of the rough contact problem requires the determination of surface displacements. The Boundary Element Method applied to rough contact problems~\cite{xu:2019} is a robust numerical workflow to obtain boundary displacements $u$ at the frontier of a deformable body, in this case a linear elastic half-space. The formulation of the continuous problem starts from the closed-form convolution of an applied traction field $p$ with a Green's function that can be expressed as the following Boundary Integral Equation (BIE): 
\begin{equation}
    u(x,y) = \frac{1-\nu^2}{\pi E}\int_{\mathcal S}
    \frac{p(\xi,\eta)}{\sqrt{(x-\xi)^2+(y-\eta)^2}}\,\mathrm d \xi d\eta.\label{eq:conv}
\end{equation}
Assuming the action of the vertical far-field displacement $\Delta$, that pushes the bodies in contact, an indentation level $\overline u$ can be defined that quantifies the interference of the rough surface and the half-space, see Figure~\ref{fig:deformation}. The introduction of this quantity  allows to state the contact problem as an infinite dimensional Linear Complementarity Problem (LCP):
\begin{align}
    w(x,y) \ge 0, \quad & p(x,y) \ge0, \notag \\
    w(x,y)p(x&,y)=0, \label{eq:lcp}
\end{align}
where $w = u -\overline u$. Equations~\eqref{eq:conv} and~\eqref{eq:lcp} above can be discretized on the Cartesian grid introduced in Section~\ref{sec_rsc}, resulting in the finite dimensional LCP:
\begin{align}
    \boldsymbol w \ge 0, \quad & \boldsymbol p \ge0, \notag\\
    \boldsymbol w \cdot \boldsymbol p =0, \quad & \boldsymbol u = \boldsymbol H \boldsymbol \cdot \boldsymbol p,\notag
\end{align}
where the discretized field variables $\boldsymbol w \in \mathbb R^{n\times n}$ and $\boldsymbol p\in \mathbb R^{n\times n}$ are coincident with the integral average value of the equivalent continuous quantities, evaluated in correspondence of the center of each cell $g_\mathrm c$. Under linear elasticity assumptions it can be proven that $\boldsymbol H $ is symmetric and positive definite and that the problem has a unique solution. The values to be used in the definition of the matrix $\boldsymbol H$ are dependent on the discretization framework employed. The interested reader is pointed to, among the others, for possible choices of the influence coefficients ~\cite{pohrt:2014,bemporad:2015}.

\begin{figure*}[thbp]
    \def\figsize{0.275}
    \def\figsizeright{0.345}
    \def\hskiplocal{\hskip 0.2cm}
    \def\rightcut{2.75cm}
    \centering
    \def\arraystretch{0.5}%
    \begin{tabular}{c@{\hskiplocal}c@{\hskiplocal}c}
    {\stackinset{c}{0.3cm}{b}{-0.45cm}{(a)}{\includegraphics[trim={0.25cm 0.25cm \rightcut cm 0.05cm},
    clip,width=\figsize\linewidth]{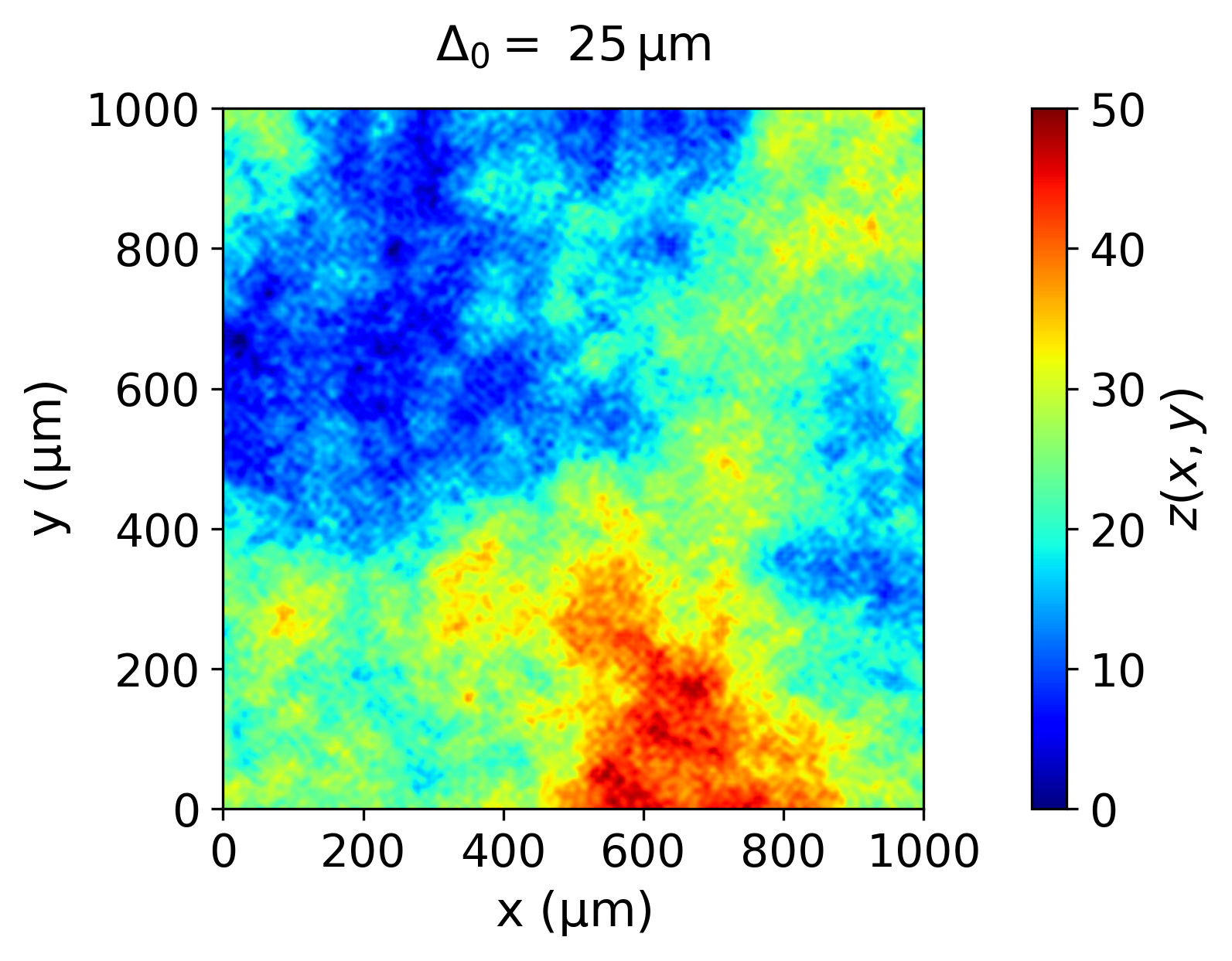}}} &
    {\stackinset{c}{0.3cm}{b}{-0.45cm}{(b)}{\includegraphics[trim={0.25cm 0.25cm \rightcut cm 0.05cm},
    clip,width=\figsize\linewidth]{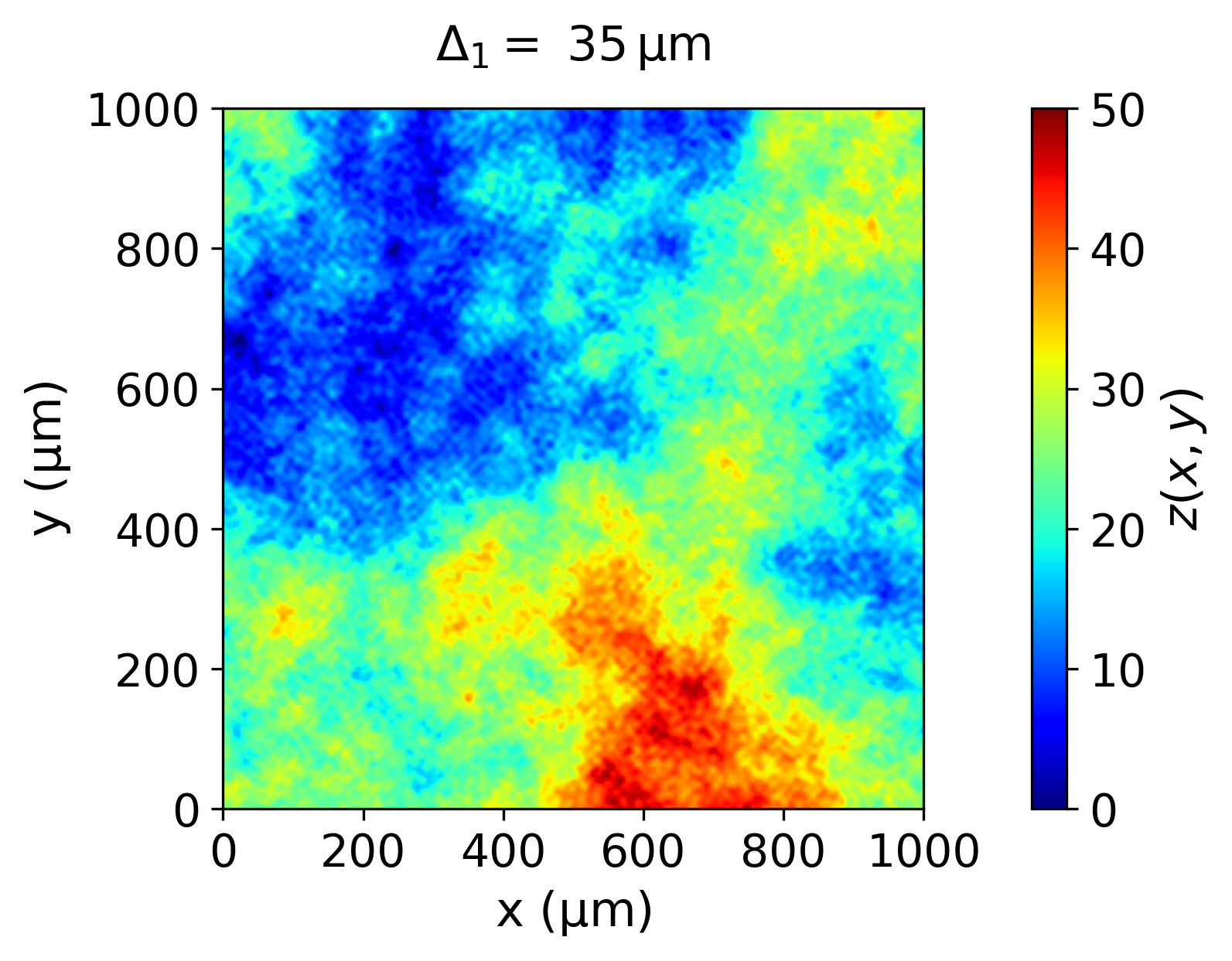}}} &
    {\stackinset{c}{-0.1cm}{b}{-0.45cm}{(c)}{\includegraphics[trim={0.25cm 0.25cm 0.1cm 0.05cm},
    clip,width=\figsizeright\linewidth]{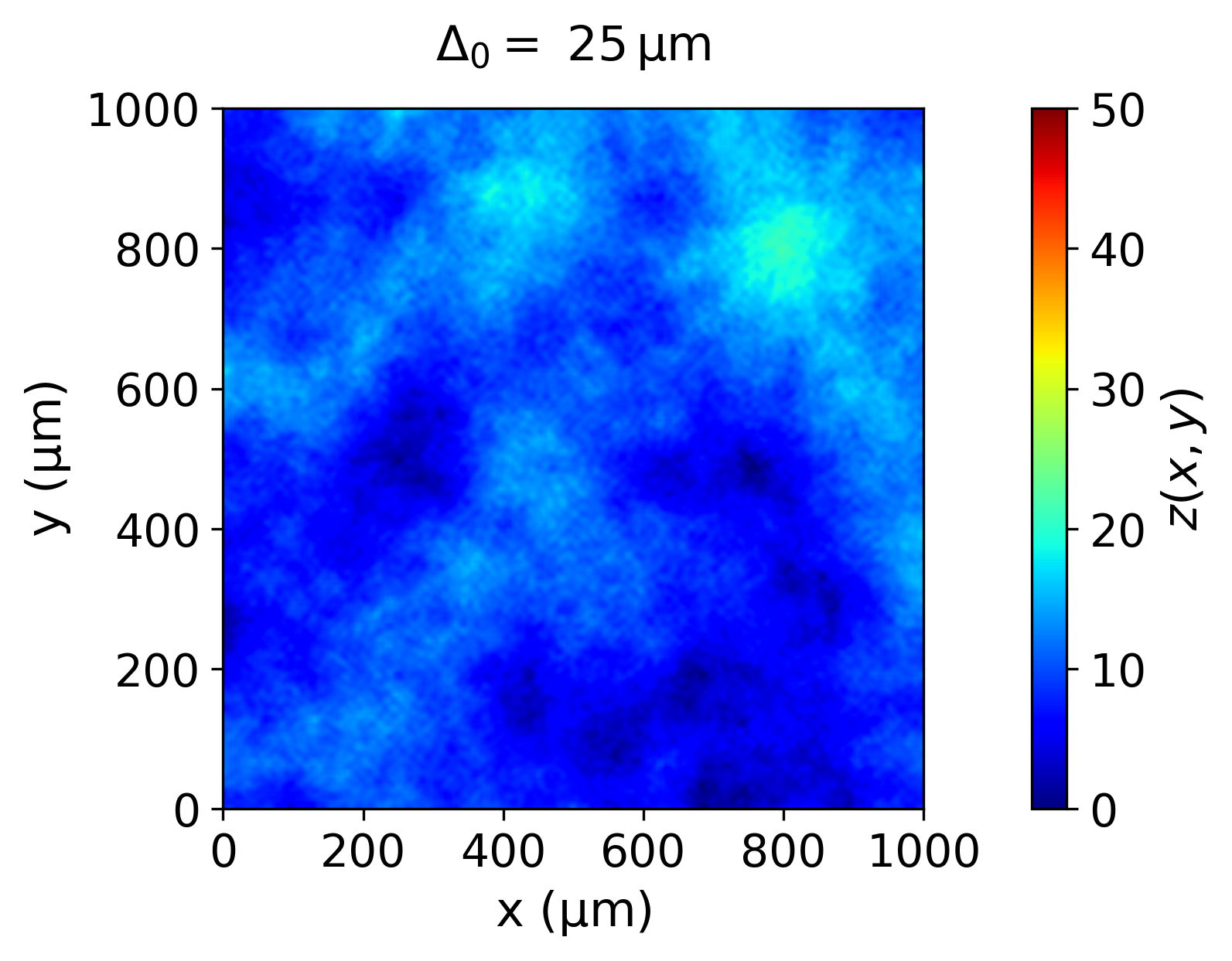}}}
    \\[4ex]
    {\stackinset{c}{0.3cm}{b}{-0.45cm}{(d)}{\includegraphics[trim={0.25cm 0.25cm \rightcut cm 0.05cm},
    clip,width=\figsize\linewidth]{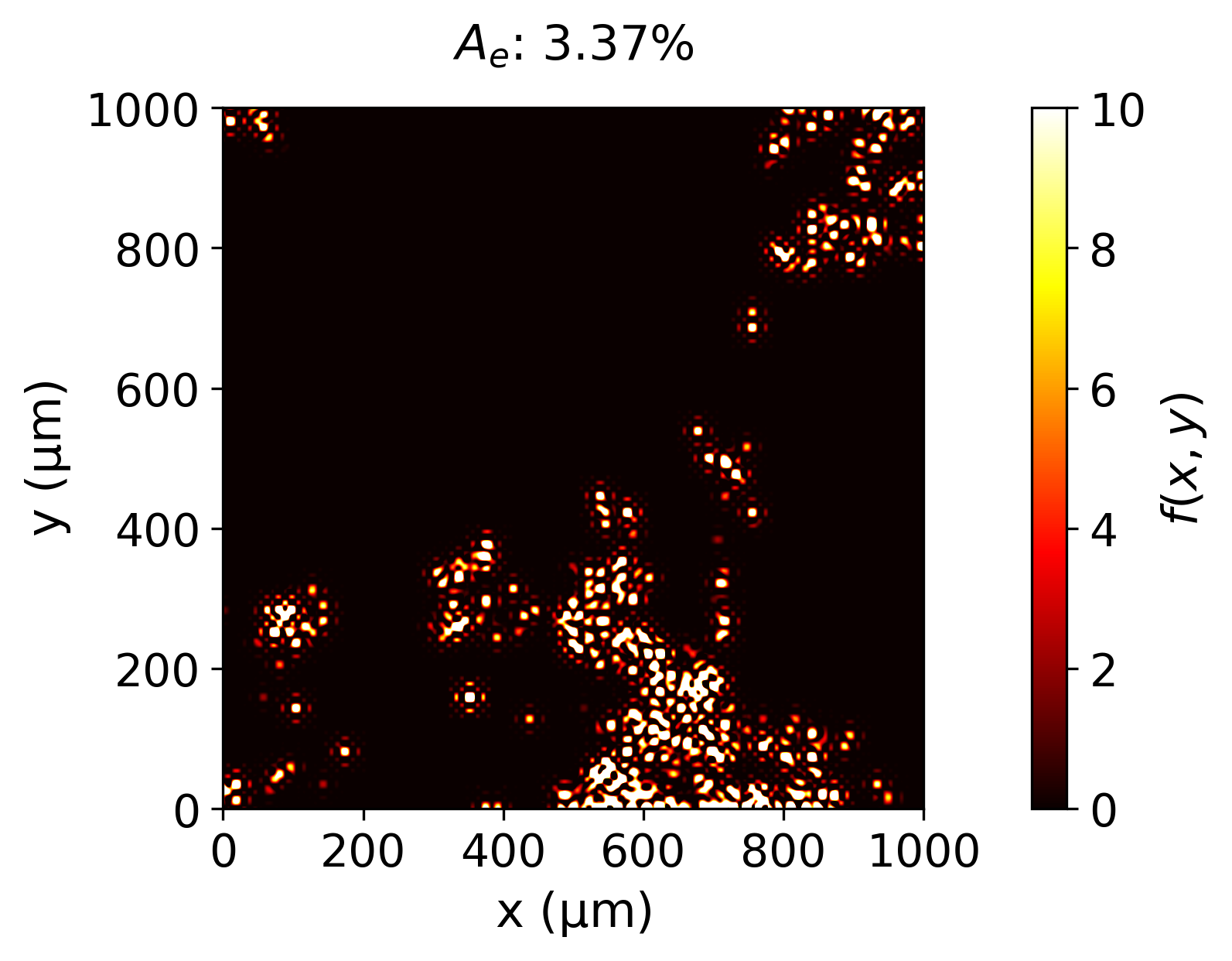}}} &
    {\stackinset{c}{0.3cm}{b}{-0.45cm}{(e)}{\includegraphics[trim={0.25cm 0.25cm \rightcut cm 0.05cm},
    clip,width=\figsize\linewidth]{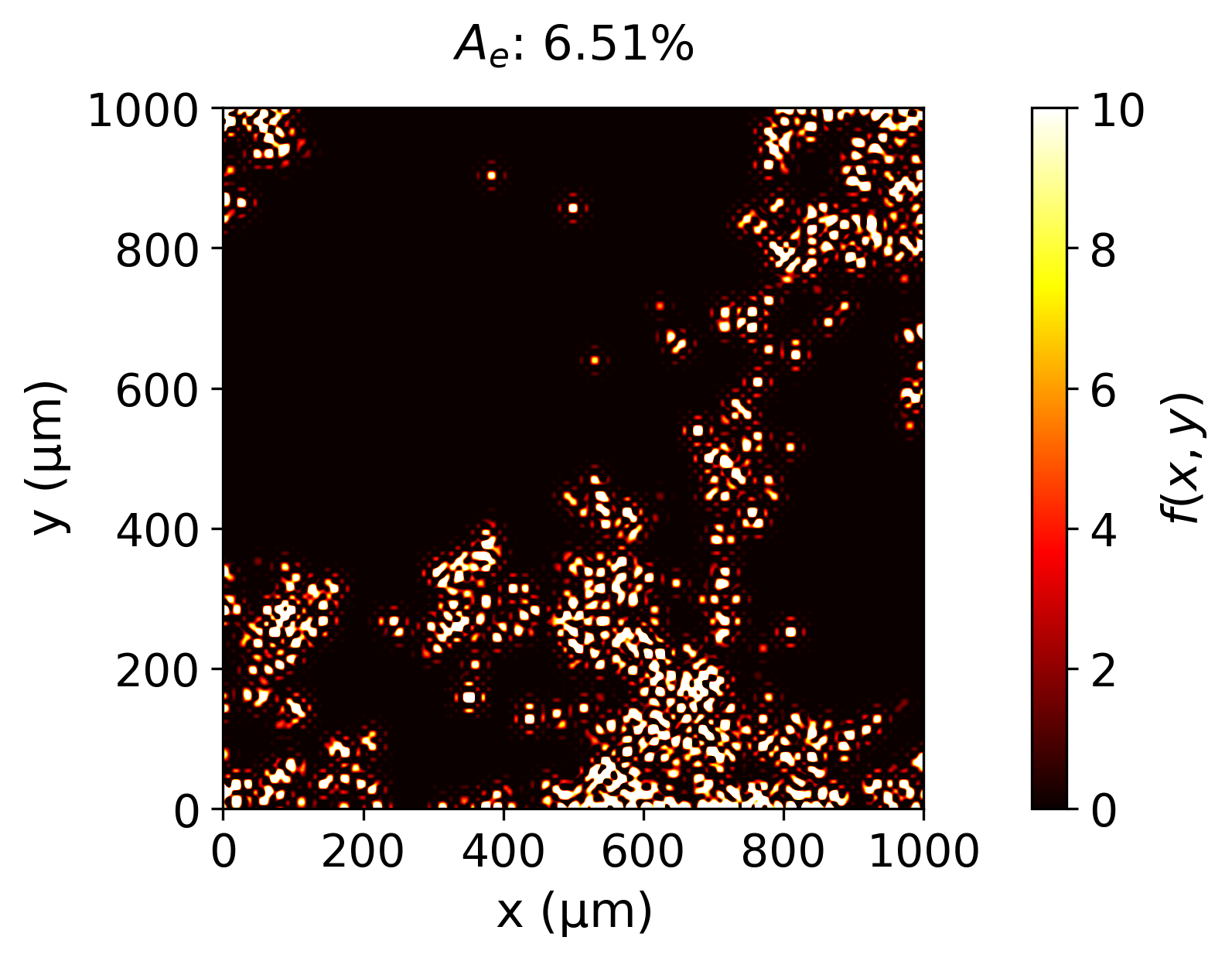}}} &
    {\stackinset{c}{-0.1cm}{b}{-0.45cm}{(f)}{\includegraphics[trim={0.25cm 0.25cm 0.1cm 0.05cm},
    clip,width=\figsizeright\linewidth]{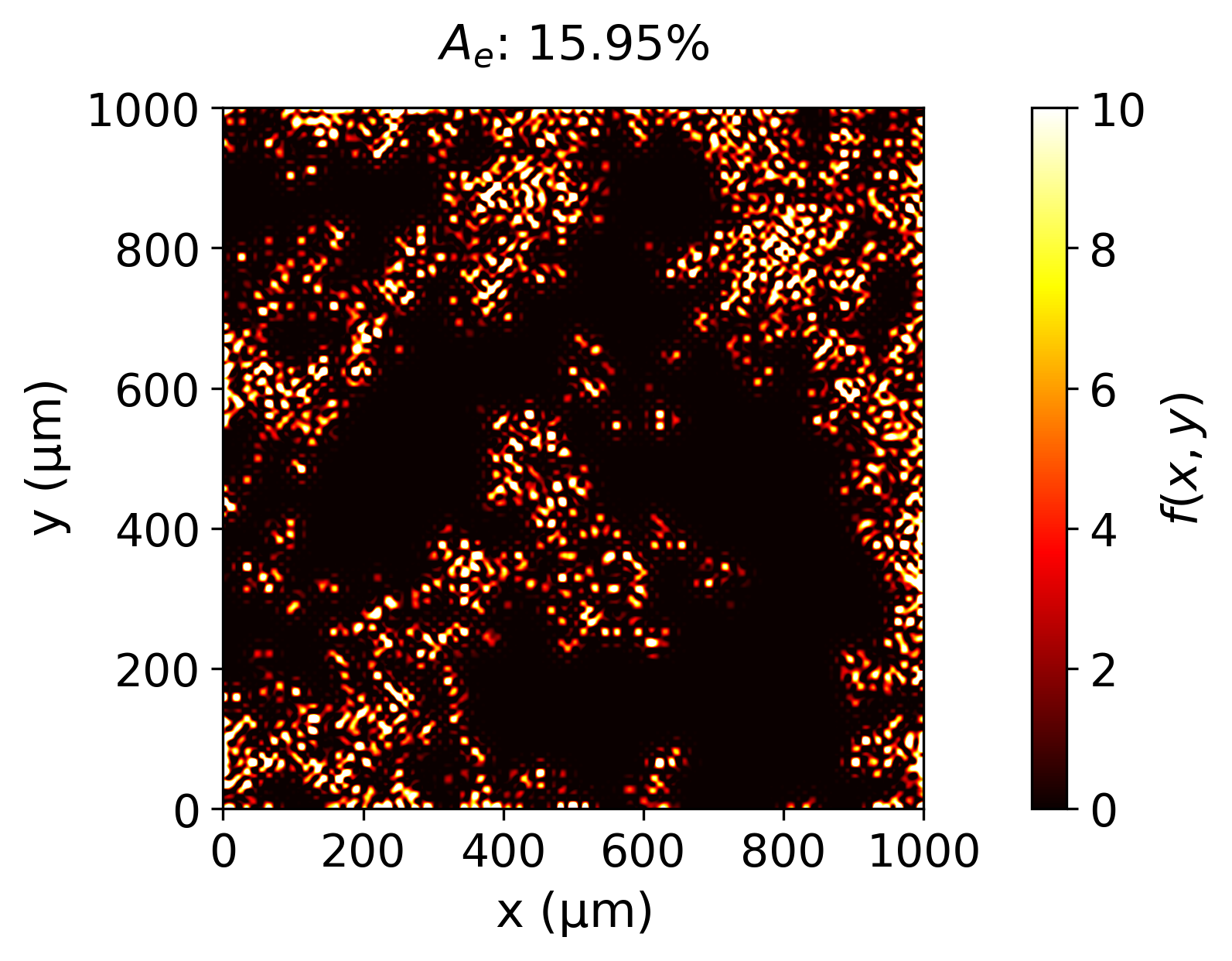}}}
    \end{tabular}
    \caption{The top row of images (Fig. a-c) shows two surface topographies generated by the RMD.  
    Dark blue indicates the deepest valleys, while dark red represents the highest summits.
    The bottom row of images (Fig. d-f) shows the resulting contact force per grid point $f$,
    for two far-field displacements \farFielDisplacementIndices{0}{25} 
    and \farFielDisplacementIndices{1}{35}.  
    The corresponding effective contact area $\EffectArea$ is also shown.} %
    \label{fig:topography_vs_pressure}
\end{figure*}

Several open-source codes are available to solve rough contact problems using the formulation presented above,
e.g. \textit{MIRCO} (co-developed by the authors themselves)~\cite{bemporad2015}, \textit{Tamaas}~\cite{frerot2020}, and \textit{MultiFEBE}~\cite{multifebe2022}. 
Since we aim at simulating linear elastic frictionless contact 
between a rigid rough indenter and an elastic half-space, \textit{MIRCO} is used to run the numerical 
simulations. 
Figure~\ref{fig:topography_vs_pressure} presents example results from contact simulations using two different surface topographies created by the RMD, and two indentation levels, \farFielDisplacementIndices{0}{25} and \farFielDisplacementIndices{1}{35}.
For the same topographies, given in Figure~\ref{fig:topography_vs_pressure}a and Figure~\ref{fig:topography_vs_pressure}b, 
increasing the indentation level leads to a larger effective contact area, as shown in 
Figure~\ref{fig:topography_vs_pressure}d and Figure~\ref{fig:topography_vs_pressure}e.
On the other hand, for the same indentation level (see Figures~\ref{fig:topography_vs_pressure}a and~\ref{fig:topography_vs_pressure}c), 
two different topographies can lead to very different  effective contact areas.
The surface in Figure~\ref{fig:topography_vs_pressure}c is much smoother and consequently more
points get in contact as can be seen from the resulting contact force distributions 
in Figures~\ref{fig:topography_vs_pressure}d and~\ref{fig:topography_vs_pressure}f.

\begin{figure*}[thbp]
    \centering
    \includegraphics[width=0.8\textwidth]{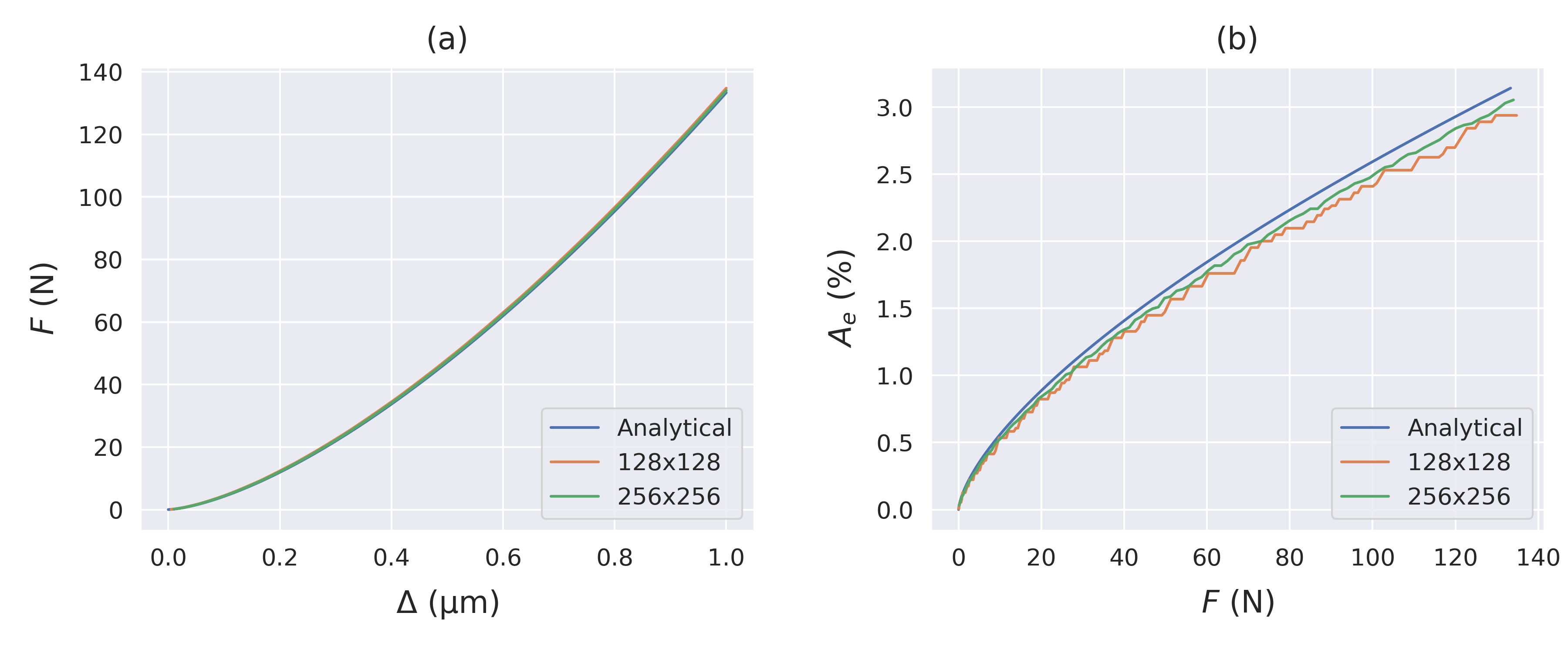}
    \caption{Comparison between analytical and BEM solutions for the paraboloid rigid indenter example. 
    BEM models have surfaces discretized on a grid with $128 \times 128$ and $256 \times 256$ points per side, respectively.}
    \label{fig:bem_vs_analy}
\end{figure*}

Furthermore, to illustrate the accuracy of \textit{MIRCO}, 
a benchmark example, namely the contact problem between an elastic half-space and a smooth paraboloid rigid indenter~\cite{popov2019}, is 
investigated. Analytical solutions for the contact force $F$ and the normalized contact area $A_n$ can be obtained as:

\begin{equation}
    F = \frac{4}{3} \frac{E}{1-\nu^2}\sqrt{R\Delta^{3}} \quad \text{and} \quad A_n = 100 \frac{\pi R \Delta}{L^2},
\end{equation}
where $R$ is the minimum radius of curvature of the paraboloid. For our specific setup, 
we set $R=\SI{5000}{\micro \meter}$, $L=\SI{1000}{\micro \meter}$, and $\Delta$ linearly varying from 0 \SI{}{\micro \meter} to 1 \SI{}{\micro \meter}. Figure~\ref{fig:bem_vs_analy} shows a comparison between the analytical solution and two numerical 
solutions obtained using BEM with two different grid discretizations of $128 \times 128$ 
and $256 \times 256$ points per side. 
As shown in Figure~\ref{fig:bem_vs_analy}a, the contact force computed using BEM matches 
the analytical solution well for both grid resolutions. 
In contrast, the computed contact area  slightly underestimates the analytical solution 
at higher resolution, and is even smaller with the coarser grid, as illustrated in Figure~\ref{fig:bem_vs_analy}b.  
These deviations stem from the method used to calculate the effective contact area. 
A simple method is used to estimate the contact area by dividing the number of contact nodes 
by the total number of nodes on the micro-scale surface. 
As a result, the finer grid leads to lower errors and smoother results.

\begin{figure*}[thbp]
    \centering
    \includegraphics[width=0.9\textwidth]{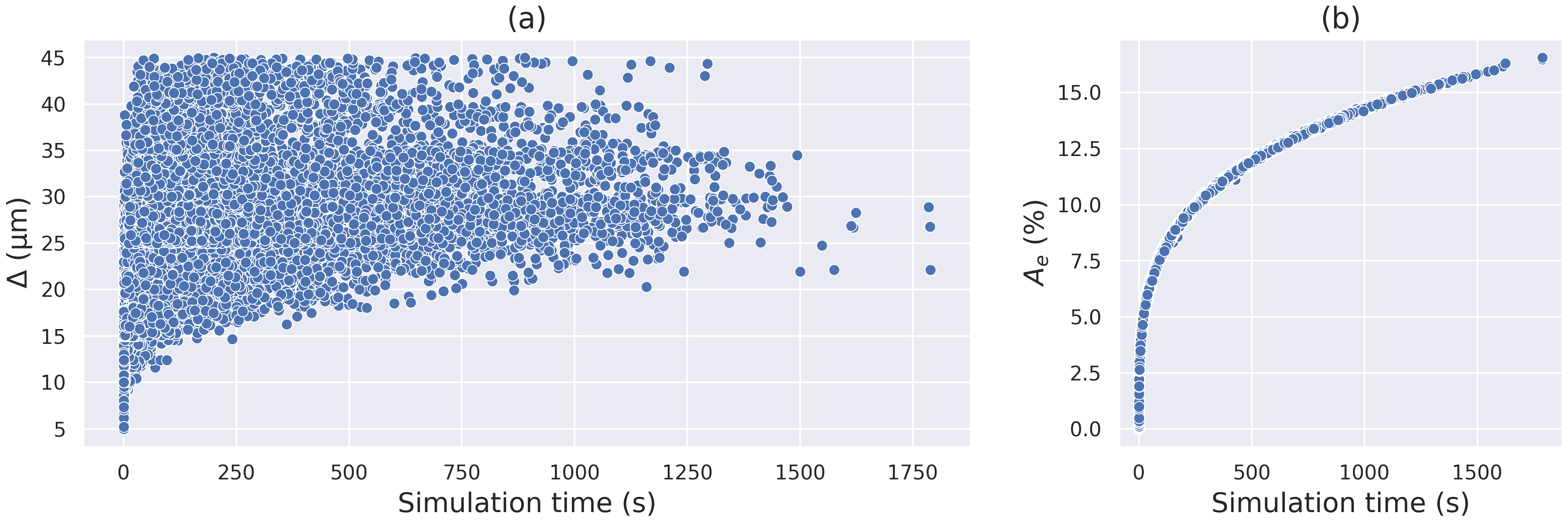}
    \caption{Comparison of simulation time for rough surface contact simulations with a grid size of $128 \times 128$.
    (a) shows the effect of different far-field displacements and surface realisations on the overall simulation time. 
    Each point represents a randomly generated surface and a specific far-field displacement.
    (b) The correlation between effective contact area and simulation time.}
    \label{fig:simulation_time}
\end{figure*}

Although BEM is an efficient method, simulations can still become computationally expensive.
Figure~\ref{fig:simulation_time}a presents the simulation times for different 
combinations of far-field displacements and surface realizations.
Each point represents the simulation time for a randomly generated surface and a specific value of the far-field displacement.
All surfaces are discretized on a $128 \times 128$ grid.
The simulation time varies significantly across different far-field displacements.
Additionally, there is considerable variability in simulation time even for the same far-field displacement, which can be attributed to differences in surface properties.
Generally, larger far-field displacements and smoother surfaces result in longer simulation times.
Notably, both higher far-field displacements and smoother surfaces are associated with increased contact areas.
Figure~\ref{fig:simulation_time}b illustrates this trend by showing the relationship between simulation time and effective contact area. Clearly, larger contact areas lead to longer simulation times.

The simulation time for a single run can reach up to approximately \SI{1750}{\second}.
While relatively low compared to other large-scale models,
this still makes multi-query analyses computationally expensive.
Such analyses---including parameter studies, sensitivity analysis, uncertainty quantification,
and parameter identification---are crucial for many real-world applications. 
For example, parameter studies and sensitivity analysis explore the influence of variables
like grid size or far-field displacement, 
while uncertainty quantification evaluates how input uncertainties affect model outputs.~\cite{Brandstaeter2021a,Wirthl2023}
Parameter identification, on the other hand, aims to infer model parameters from data.~\cite{willmann2022-BayesianCalibrationCoupled,Dinkel2024}
All these tasks require running the model many times---often several thousands---making them potentially prohibitively expensive due to the computational cost of individual BEM simulations.
Similarly, multi-scale algorithms---where different models are used to resolve different scales of a problem, with information exchanged between scales---can become prohibitively expensive even for moderately sized problems due to their nested simulation structure.
For instance, a micro-scale BEM simulation may be required at each Gauss point of a continuum FEM simulation.~\cite{bonari:2020b}
To alleviate this computational burden, data-driven surrogate models can be employed to replace the costly micro-scale simulations, thereby making large-scale multi-scale analyses significantly more tractable and computationally efficient.

\subsection{Data-driven problem: learning a surrogate model}\label{sec_surrogate_modeling}
Surrogate modeling aims to reduce computational costs by approximating complex, computationally expensive
models with more efficient alternatives.
Often, this reduction in computational cost comes at the expense of model accuracy and generality.
Surrogate models can take various forms, such as analytical models~\cite{danwitz2023}, physics-informed machine learning models~\cite{sahin2024b}, or purely data-driven approaches based on machine learning techniques~\cite{ganti2020}. 
The choice of surrogate model depends on the desired balance between 
model accuracy, generality, and computational efficiency.

Mathematically, the data-driven surrogate modeling task can be formulated as a regression task.
Given a dataset (or database) $\mathcal{D}=\{(\vec{x}_i, y_i)\}_{i=1,\ldots,N}$ of $N$ pairs of inputs $\vec{x}_i$ and
corresponding outputs $y_i=f(\vec{x}_i)$,  the goal is to learn a function  
$\SurrogateModel:\mathbb{R}^d \rightarrow \mathbb{R}; \vec{x} \mapsto \hat{y} = \SurrogateModel(\vec{x}) $
 that approximates the true model 
$\Model:\mathbb{R}^d \rightarrow \mathbb{R};  \vec{x} \mapsto y = \Model(\vec{x})$ 
such that 
\begin{equation}
   \SurrogateModel \approx \Model ,
\end{equation}
where $\vec{x}\in\mathbb{R}^d$ is the $d$-dimensional vector of input parameters, $y\in\mathbb{R}$ is the scalar quantity of interest (or target quantity), $\hat{y}\in\mathbb{R}$ is the approximation of the quantity of interest by the surrogate model, and $\SurrogateParameters$ are the free parameters of the surrogate model.
Learning the surrogate model $\SurrogateModel$ typically involves minimizing a loss function 

\begin{equation}
    \mathcal{L}(\mathcal{D}, \SurrogateModel) = \sum_{i=1}^N (y_i - \hat{y}_i)^2,
\end{equation}
with respect to the free parameters $\SurrogateParameters$, where $\hat{y}_i = \SurrogateModel(\vec{x}_i)$ denotes the surrogate prediction at sample point $\vec{x}_i$.

Given this mathematical framework, we propose a data-driven surrogate modeling workflow to reduce the cost of evaluating the effective contact area in rough surface contact problems, which depend on the far-field displacement $\Delta$ and the surface height field $\HeightField$, as defined in Section~\ref{sec_rsc}.  
In this context, the system is described by the function $f$ introduced in Equation~\eqref{eq:true_model} and illustrated in Figure~\ref{subfig:physics}.  
The input vector is defined as $\vec{x} = [\Delta, \HeightField]$, and the target quantity is $y = \mathcal{A}_\mathrm{e}$.  
We aim to approximate $f$ using a surrogate model $\SurrogateModel$, as depicted in Figure~\ref{subfig:surrogate}.  

A central challenge in this problem is the high dimensionality $d$ of the input vector, primarily due to the discretized height field $\HeightField$.  
To address this, we adopt a dimensionality reduction approach: instead of using $\boldsymbol{z}$ directly, we represent it through a set of summary statistics or statistical descriptors $\boldsymbol \vartheta$ (cf.~Section~\ref{sec_stat_of_top}).  
This significantly reduces the input dimensionality, as $\Statistics$ captures the essential characteristics of $\HeightField$ with far fewer parameters.  

Moreover, while the dimensionality of $\HeightField$ increases with the grid size (e.g., 1024 for a $32 \times 32$ grid and 16384 for a $128 \times 128$ grid), the number of statistical parameters in $\Statistics$ remains fixed.  
In this study, we compute 22 such statistical parameters, enabling efficient and scalable surrogate modeling.

Ultimately, the surrogate model $\SurrogateModel$ can be equivalently described by the following set of equations:

\begin{equation}
    \hat{\mathcal A_\mathrm e} = \SurrogateModel(\Delta, \boldsymbol z) \iff
        \begin{cases}
            \Statistics &= g(\HeightField),\\
            \hat{\mathcal A_\mathrm e} &= \hat{h}_{\SurrogateParameters}(\Delta, \Statistics),
        \end{cases}
        \label{eq:surrogate}
\end{equation}
Here, $g$ is a known deterministic function that maps the height field $\HeightField$ to a vector of statistical descriptors $\Statistics$ (as detailed in Section~\ref{sec_stat_of_top}), and $\hat{h}_{\SurrogateParameters}$ is a regression model trained from data. 
This model is learned using the dataset $\mathcal{D} = \left\{ ( [\Delta_i, \Statistics_i], \mathcal A_\mathrm{e}^i ) \right\}_{i=1,\ldots,N}$, where each $\Statistics_i = g(\boldsymbol{z}_i)$.  
    
A graphical representation of Equation~\eqref{eq:surrogate} is shown in Figure~\ref{subfig:surrogate}.  
Once the model $\hat{h}_{\SurrogateParameters}$ is trained, evaluating the surrogate on a new, unseen rough surface involves computing its statistical descriptors via $g$ and applying the learned regression model as outlined in Equation~\eqref{eq:surrogate}.
While the original model $f$ could also be used to predict the target quantity, doing so would incur significant computational cost.  
In contrast, evaluating the surrogate model is substantially faster, making it particularly well-suited for multi-query analysis scenarios (cf.~previous section), where thousands of such model evaluations may be required.

\section{Methodolody} \label{sec_methods}
\subsection{Surrogate Modeling Workflow}\label{sec_workflow}
\begin{figure*}[thbp]
    \centering
    \includegraphics[width=0.85\textwidth]{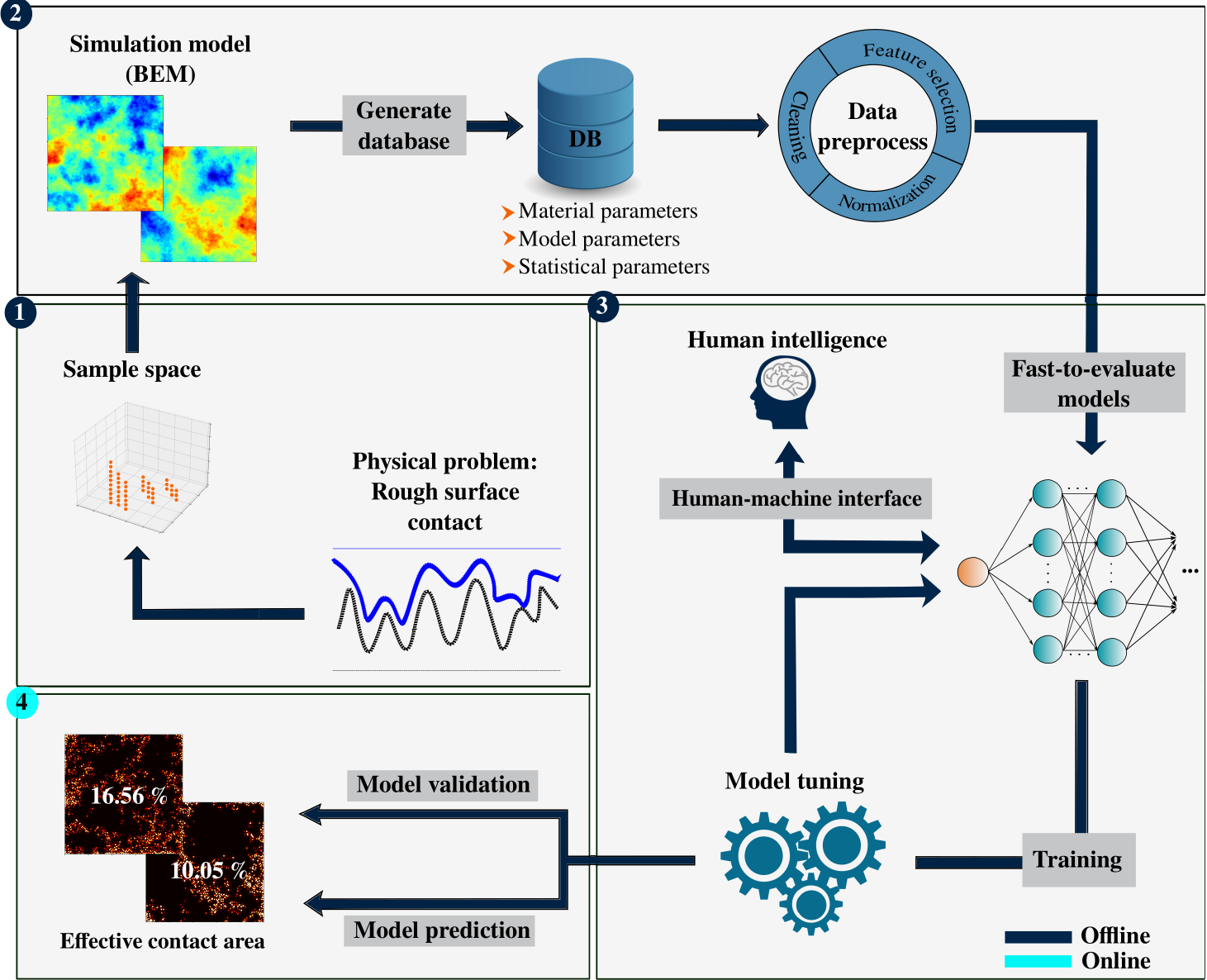}
    \caption{Data-driven surrogate modeling workflow to solve rough surface contact simulations.}
    \label{fig:surrogate_modeling_workflow}
\end{figure*}

After mathematically defining the problem setup in the previous section, we describe the full modeling workflow from a practical perspective.  
The proposed modeling workflow is illustrated in Figure~\ref{fig:surrogate_modeling_workflow}.
First, we define the input parameter space and sample from it.
The input parameter space consists of the statistical parameters of a rough surface and the far-field displacement.  
We use controlled sampling~\cite{goodman1950} to explore this space.  
While the far-field displacement can be sampled directly, the surface parameters are sampled indirectly:
a random rough surface is first generated using the RMD algorithm (see Section~\ref{sec_rmd}),
and its statistical parameters are then computed as described in Section~\ref{sec_stat_of_top}.  

Second, we run BEM simulations for all input samples (pairs of rough surfaces and far-field displacements) to
compute the corresponding effective contact areas.  
In this way, we construct a database $\mathcal{D}$ consisting of input-output samples.  
An important practical aspect of data-driven surrogate modeling is data preprocessing, 
which follows database generation.
Not all parameters may be relevant, and some data points might be incomplete, e.g., due to failed BEM 
simulations.  
Thus, methods such as feature selection and data cleaning are applied to obtain the final form of the 
database~\cite{geron2022hands}. 
Feature selection is the process of identifying and selecting a subset of the most relevant input parameters 
(features) from the dataset that contribute the most to the model's predictive performance.  
The goal is to improve model accuracy, reduce overfitting, and decrease training time by removing irrelevant,
redundant, or noisy features.  
Data cleaning is used to remove missing values resulting from failed simulations. 
Moreover, data normalization~\cite{geron2022hands, singh2020} is applied to scale the problem and improve model performance and stability.

Third, various machine learning regression models are trained as fast-to-evaluate surrogate models, and hyperparameter optimization is performed to tune them further.  
Due to the \textit{black-box} nature of purely data-driven methods, human oversight is necessary throughout the training process.  
This includes defining the hyperparameter search space, managing the workflow, making critical decisions, and selecting the optimal model.  
In this work, the performance of the trained surrogate models is evaluated on unseen but known test data (a subset of the database) using accuracy metrics, such as the mean square error (see Section~\ref{sec_accuracy_metrics}).  

Finally, in the fourth step, the surrogate model can be used to predict the quantity of interest for any new input point, for example during a multi-query analysis such as a parameter study 
or uncertainty quantification.  
This stage is referred to as the online phase.  
The first three steps (cf.~Figure~\ref{fig:surrogate_modeling_workflow}) take place in the offline phase, meaning they are performed only once.  

The workflow is implemented in our in-house package called \textit{DeepSim} 
(\url{https://github.com/imcs-compsim/deep_sim}), which is developed to automate numerical simulations,
and machine learning frameworks.
{\color{red}{(The code will be available after acceptance)}}
\subsection{Generation of Database}\label{sec_data_generation}
\begin{figure*}[tbhp]
    \centering
    \includegraphics[width=1\textwidth]{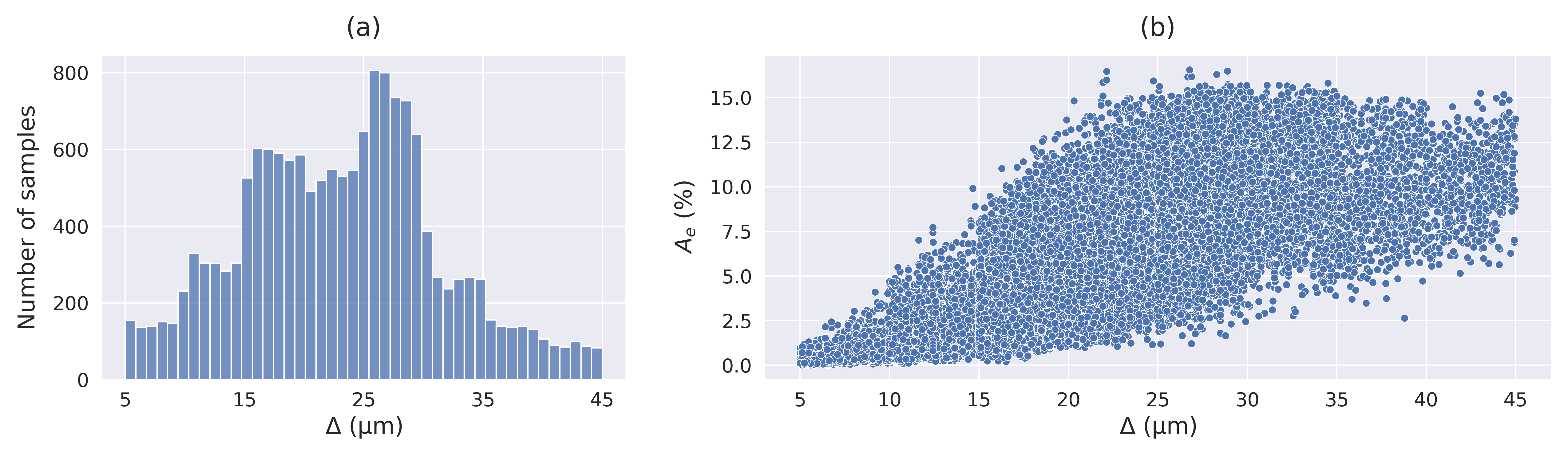}
    \caption{Visualization of the database.
    (a) distribution of far-field displacement, and 
    (b) scatter plot showing the target effective contact area against the feature far-field displacement.
    Each point represents a triplet consisting of a rough surface,
    a far-field displacement, and the corresponding effective contact area.
    Variations in effective contact area at a fixed far-field displacement are due to differences in the characteristics of the rough surfaces.}
    \label{fig:database}
\end{figure*}

To generate the database, we first sample from the input parameter space and then run BEM simulations for each sample.  
The input space includes the statistical parameters of rough surfaces and the far-field displacement. 
As described in Section~\ref{sec_rsc}, the material properties do not influence the contact pattern.  
Therefore, parameters such as Young's modulus and Poisson's ratio are held constant throughout this study.  
Additionally, more samples are placed in regions of greater interest through controlled sampling~\cite{goodman1950}.
The far-field displacement is sampled directly from the range \SI{5}{\micro\meter} to \SI{45}{\micro\meter}, with a higher concentration of samples for $\Delta \leq \SI{25}{\micro\meter}$ (see Figure~\ref{fig:database}a and Figure~\ref{fig:database}b), since larger far-field displacements lead to prohibitive computational cost (cf.~Figure~\ref{fig:simulation_time}).  
Statistical surface parameters are sampled indirectly: rough surfaces with a $128 \times 128$ grid are generated using the RMD algorithm (see Section~\ref{sec_rmd}), and their statistical properties are computed as described in Section~\ref{sec_stat_of_top}.  
BEM simulations are performed for each input sample to compute the corresponding effective contact area.

The resulting input-output pairs form the desired database.  
In total, we run 15,878 BEM simulations, with an average simulation time of \SITime{190} per model evaluation.  
A comprehensive list of all parameters used in this study is provided in Table~\ref{tab:list_of_params}.  
The table includes fixed parameters (e.g., material and model parameters) used to generate the database, along with the observed ranges for all input and target quantities.  
The input features for the data-driven surrogate models consist of the far-field displacement and 22 statistical parameters describing the rough surfaces, as listed in Table~\ref{tab:list_of_params}.  
The database includes one quantity of interest, the effective contact area,
which serves as the model output.  

Note that the table presents raw data, before preprocessing (see Section~\ref{sec_workflow}).  
In particular, an $L_2$-normalization is applied to ensure each feature vector has unit length:

\begin{equation}
    X_i = \frac{x_i}{\sqrt{\sum_{i=1}^{n_f} x_i^2}}, 
\end{equation}
where $x_i$ is the non-normalized feature, $n_f$ is the number of features, and $X_i$ is the normalized feature. 

After preprocessing, the database is used to train fast-to-evaluate surrogate models.  
We adopt an 80:20 train-test split, resulting in a training set with 12,703 samples and a test set with 3,175 samples.  
Each sample has 23 features and one scalar target.

\begin{table*}[thbp]
    \centering
    \caption{The list of parameters of the generated raw database including material parameters,
    model parameters, statistical parameters, and the target quantity. Each row contains a parameter with
    corresponding symbol and the range of variation.}
    \label{tab:list_of_params}
    \begin{tabular}{clcc}
    \hline 
    \textbf{Type} 
    & \textbf{Name}
    & \textbf{Symbol} 
    & \textbf{Range} 
    \\ \hline 
    \multirow{3}{*}{\begin{tabular}[c]{c}Model \\ parameters \end{tabular} } 
    & Scan length (\displacementUnit) & $L$ & \num{1000} \\
    &Hurst exponent (-) & $H$ & $[0.5,0.8]$ \\
    & Far-field displacement (\displacementUnit) & $\Delta$ & $[5,45]$ 
    \\ \hline 
    \multirow{22}{*}{\begin{tabular}[c]{c}Statistical \\ parameters \end{tabular} }
    & Mean of peaks (\displacementUnit)             & $\bar{z}_{\text{p}}$ & $[8.20,47.51]$ \\
    & Root mean square of peaks (\displacementUnit) & $\text{RMS}[{z}_{\text{p}}]$ & $[2.54,15.49]$ \\
    & Kurtosis of peaks (-)               & $\text{K}[{z}_{\text{p}}]$ & $[1.56,7.13]$ \\
    & Skewness of peaks (-)               & $\text{Sk}[{z}_{\text{p}}]$ & $[-1.34,1.36]$ \\
    & Density of peaks (-)                & $\rho_{\text{p}}$ & $[0.19,0.27]$\\
    & Mean of curvature of peaks (\InversedisplacementUnit) & $\bar{\kappa}_{\text{p}}$        & $[0.015,0.066]$\\
    & Kurtosis of curvature of peaks (-)     & $\text{K}[{\kappa}_{\text{p}}]$  & $[-0.07,5.35]$ \\
    & Skewness of curvature of peaks (-)     & $\text{Sk}[{\kappa}_{\text{p}}]$ & $[0.55,1.15]$ \\
    & Bandwidth parameter in x direction (-) & $\alpha_x$                           & $[4.06,46.19]$ \\
    & Bandwidth parameter in y direction (-) & $\alpha_y$                           & $[4.19,48.52]$ \\
    & Mean of asperities (\displacementUnit)              & $\bar{z}_{\text{a}}$         & $[8.54,48.18]$ \\
    & Root mean square of asperities (\displacementUnit)  & $\text{RMS}[{z}_{\text{a}}]$ & $[2.56,15.60]$ \\
    & Kurtosis of asperities (-)                & $\text{K}[{z}_{\text{a}}]$   & $[-1.45,4.26]$ \\
    & Skewness of asperities (-)                & $\text{Sk}[{z}_{\text{a}}]$  & $[-1.34,1.48]$ \\
    & Density of asperities (-)                 & $\rho_{\text{a}}$            & $[0.08,0.13]$\\
    & Mean of curvature of asperities (\InversedisplacementUnit)              & $\bar{\kappa}_{\text{a}}$         & $[0.017,0.072]$ \\
    & Root mean square of curvature of asperities (\InversedisplacementUnit)  & $\text{RMS}[{\kappa}_{\text{a}}]$ & $[0.007,0.031]$ \\
    & Kurtosis of curvature of asperities (-)                  & $\text{K}[{\kappa}_{\text{a}}]$   & $[2.81,15.14]$ \\
    & Skewness of curvature of asperities (-)                  & $\text{Sk}[{\kappa}_{\text{a}}]$  & $[0.41,1.6]$ \\
    & Mean of surface height (\displacementUnit)              & $\bar{z}$        & $[7.71,45.36]$ \\
    & Max. of surface height (\displacementUnit)              & $z^{\text{max}}$ & $[16.36,79.25]$ \\
    & Root mean square of surface height (\displacementUnit)  & $\text{RMS}[z]$  & $[2.62,15.55]$
    \\ \hline
    \begin{tabular}[c]{c}Target \\ parameter \end{tabular}
    & Effective contact area (\%) 
    & $A_e$
    & $[0.01,16.56]$
    \\ \hline
    \end{tabular}
\end{table*} 
\subsection{Fast-to-Evaluate Modeling with Hyperparameter Optimization}\label{sec_data_driven_modeling}
To find the optimal data-driven model, 
we compare different regression models, including Decision Tree~\cite{Wu2008}, 
Random Forest~\cite{Breiman2001}, Support Vector Machine (SVM) Regressor~\cite{vapnik95}, 
Adaptive Boosting (AdaBoost)~\cite{freund1997}, Gaussian Process (GP) Regressor~\cite{rasmussen2005}, 
Gradient Boost~\cite{friedman2001}, Kernel Ridge~\cite{Vovk2013}, 
eXtreme Gradient Boosting (XGBoost)~\cite{chen2016}, and Multi-Layer Perceptron (MLP) Regressor~\cite{goodfellow2016}. 

\begin{figure*}[tbhp]
    \centering
    \includegraphics[width=\textwidth]{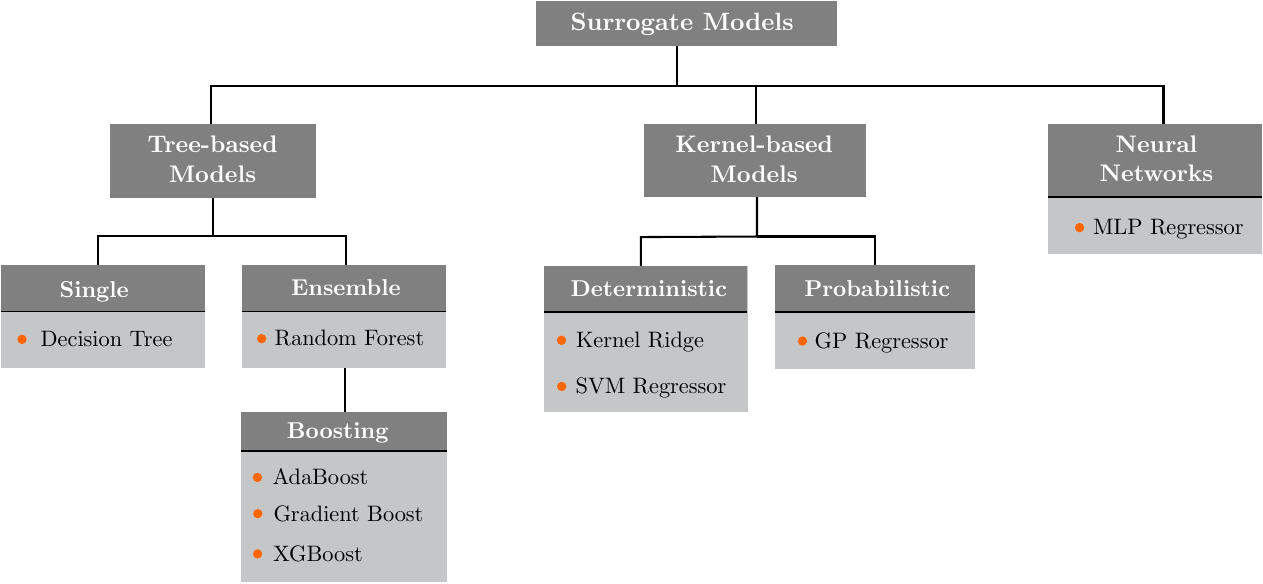}
    \caption{Classification of surrogate modeling techniques based on their characteristics.}
    \label{fig:models}
\end{figure*}

Figure \ref{fig:models} illustrates the classification of all the deployed machine learning algorithms based on their 
characteristics, categorized into three main groups: tree-based models, kernel-based models, and neural networks.

Tree-based models are further divided into single and ensemble methods, depending on whether the 
model is a standalone decision tree or it is a combination of multiple decision trees. Boosting algorithms, on the other 
hand, sequentially construct ensemble algorithms to enhance prediction accuracy.  

Kernel-based regression methods are non-parametric techniques that use kernel functions to map input data into high-dimensional feature spaces, allowing them to capture complex, nonlinear relationships.  
They leverage the \textit{kernel trick} to compute inner products in the transformed space without explicitly performing the transformation~\cite{Murphy2011}. 
Among the kernel-based models used, Kernel Ridge and SVM Regressors are deterministic, while the GP Regressor is probabilistic.  
When using the same kernel and hyperparameters, the mean prediction of a GP Regressor is equivalent to that of Kernel Ridge Regression (see Appendix~\ref{Appendix1}).  
The SVM Regressor aims to find a function (or hyperplane) that deviates from the actual target values by at most a specified margin (epsilon) for as many data points as possible, while also maintaining model simplicity (i.e., minimizing the norm of the weights).

The MLP Regressor, in contrast, is a neural network-based method capable of modeling nonlinear relationships through learned weights and activations.
Note that the MLP Regressor with regularization and a single-layer architecture using a linear activation function and the Kernel Ridge algorithm employing a linear kernel exhibit similar characteristics.  

Many machine learning models have hyperparameters.
Hyperparameters are configuration settings that are set before training begins.
They control the learning process (e.g., learning rate, batch size, number of layers) and significantly influence model performance but are not learned from the data.
Since the choice of hyperparameters directly affects model accuracy and performance~\cite{yang2020}, we perform a hyperparameter optimization using grid search with cross-validation to identify the best-performing configuration of a model~\cite{Feurer2019}.
By comparing the best-performing configuration of each model, we ensure a fair comparison across models.  

In a grid search, the model is trained on all combinations of hyperparameters on a predefined grid:
First, a search space is specified for each hyperparameter. Second, the search space is sampled on a fixed number of grid points.
Grid-based hyperparameter optimization suffers from the curse of dimensionality and can quickly become computationally very expensive.  
Despite its computational cost, we demonstrate that the effort is justified, as it leads to substantially improved model performance (cf.~Section~\ref{sec_hyper}).
To balance computational cost, we focus on tuning key hyperparameters only~\cite{yang2020}. 
For example, in tree-based algorithms such as XGBoost, important hyperparameters include the learning rate, maximum tree depth, and the number of estimators, corresponding to the maximum number of boosting trees~\cite{gorishniy2021}. 
A full list of the investigated hyperparameters is provided in Table~\ref{tab:hyperparameters}.  
Detailed descriptions of the machine learning models and their corresponding hyperparameters can be found in Pedregosa et al.~\cite{scikitlearn2011}.

To avoid setting aside a portion of the training data exclusively for validation (note that we already split the database into 80\% training and 20\% test data, cf.~Section~\ref{sec_data_generation}), we use $k$-fold cross-validation during hyperparameter optimization.  
This method splits the training data into $k$ equal parts, or folds.  
The model is then trained $k$ times, each time using $k-1$ folds for training and the remaining fold for validation.  
This ensures that every data point is used for both training and validation, providing a more reliable estimate of model performance.  
The final performance metric is obtained by averaging the results over all $k$ runs.  
For more details on $k$-fold cross-validation, we refer readers to the original literature~\cite{kohavi1995,zhang2023}. 
In this study, we specifically apply 5-fold cross-validation. 

\begin{table*}[thbp]
    \centering
    \caption{Investigated surrogate model hyperparameter optimization, including the model type, selected hyperparameters,
    the defined parameter search space, and the optimal parameter value.}
    \label{tab:hyperparameters}
    \begin{tabular}{llccc}
    \hline
    \textbf{Model} 
    & \multicolumn{1}{c}{\textbf{Hyperparams.}}                                                     
    & \textbf{Search space}                                   
    & \textbf{Grid points}                                   
    & \textbf{Optimum}                                   
    \\ \hline
    \begin{tabular}[c]{@{}c@{}}Decision \\ Tree \end{tabular}     
    & \begin{tabular}[c]{@{}l@{}}max depth \\ max features\\ min samples leaf\\ min samples split\end{tabular}      
    & \begin{tabular}[c]{@{}c@{}}1 - 30\\ auto, sqrt, log2, none \\ 1 - 15\\ 2 - 10\end{tabular}                            
    & \begin{tabular}[c]{@{}c@{}}5\\ 4\\ 6\\ 5\end{tabular}      
    & \begin{tabular}[c]{@{}c@{}}30\\ none\\ 7\\ 10\end{tabular}                  
    \\ \hline
    \begin{tabular}[c]{@{}c@{}}Random \\ Forest \end{tabular}  
    & \begin{tabular}[c]{@{}l@{}}max depth\\ max features\\ \# estimators \\criterion \\min samples split \\min samples leaf \end{tabular}                                
    & \begin{tabular}[c]{@{}c@{}}5 - 100\\ sqrt, log2, 0.3, 1 \\20 - \num{2000} \\sq. err., abs. err., Friedman, Poisson \\ 2 - 10 \\ 1 - 15\end{tabular}
    & \begin{tabular}[c]{@{}c@{}}9\\ 4 \\ 11 \\ 4 \\ 3 \\ 4\end{tabular}         
    & \begin{tabular}[c]{@{}c@{}}{20}\\ {1} \\ {\num{1000}} \\ {sq. err.} \\ {2} \\ {1} \end{tabular}                      
    \\ \hline
    AdaBoost        
    & \begin{tabular}[c]{@{}l@{}}max depth\\  loss\\ \# estimators \\ learning rate \end{tabular}         
    & \begin{tabular}[c]{@{}c@{}}3 - 35\\  linear, square, exponential\\ 10 - 1500 \\ \num{1e-2} - 2.0 \end{tabular} 
    & \begin{tabular}[c]{@{}c@{}}7\\ 3\\ 11 \\ 9\end{tabular} 
    & \begin{tabular}[c]{@{}c@{}}18\\ square\\ \num{1500} \\ 1.0 \end{tabular}       
    \\ \hline
    \begin{tabular}[c]{@{}c@{}}Gradient \\ Boost \end{tabular}  
    & \begin{tabular}[c]{@{}l@{}}learning rate \\loss \\ max depth\\ \# estimators \\ criterion \\ max features \end{tabular} 
    & \begin{tabular}[c]{@{}c@{}}\num{1e-3} - 2.0\\ sq. err., abs. err., huber, quantile\\ 5 - 100\\ 50 - \num{1600} \\ Friedman, sq. err. \\ auto, sqrt, log2 \end{tabular}          
    & \begin{tabular}[c]{@{}c@{}}8\\ 4\\ 8\\ 10 \\ 2 \\ 3\end{tabular} 
    & \begin{tabular}[c]{@{}c@{}}0.05\\ sq. err. \\ 5\\ \num{1600} \\Friedman \\auto \end{tabular}         
    \\ \hline
    XGBoost         
    & \begin{tabular}[c]{@{}l@{}}learning rate\\ max depth\\ \# estimators\\ gamma\end{tabular}        
    & \begin{tabular}[c]{@{}c@{}}\num{1e-2} - 2.0\\ 3 - 15\\ 10 - \num{1500}\\ 0 - 2\end{tabular}                    
    & \begin{tabular}[c]{@{}c@{}}8\\ 5\\ 14\\ 8\end{tabular} 
    & \begin{tabular}[c]{@{}c@{}}0.1\\ 3\\ \num{1500}\\ 0.0\end{tabular}            
    \\ \hline
    \begin{tabular}[c]{@{}c@{}}SVM \\ Regressor \end{tabular}    
    & \begin{tabular}[c]{@{}l@{}}C\\ degree \\ kernel \\gamma \\cache size \end{tabular}                                                
    & \begin{tabular}[c]{@{}c@{}}0.1 - \num{1400}\\ 3 - 7 \\ linear, rbf, poly, sigmoid\\ scale, auto \\ 200 - 5000\end{tabular}                                      
    & \begin{tabular}[c]{@{}c@{}}12\\ 3 \\ 4 \\ 2 \\ 6\end{tabular}       
    & \begin{tabular}[c]{@{}c@{}}\num{1400}\\ 7 \\ poly \\ scale \\200 \end{tabular}                      
    \\ \hline
    \begin{tabular}[c]{@{}c@{}}GP \\ Regressor \end{tabular}         
    & \begin{tabular}[c]{@{}l@{}} alpha \end{tabular}         
    & \begin{tabular}[c]{@{}c@{}}\num{1e-3} - 1.0 \end{tabular} 
    & \begin{tabular}[c]{@{}c@{}} 110 \end{tabular} 
    & \begin{tabular}[c]{@{}c@{}} 0.0374 \end{tabular}       
    \\ \hline
    \begin{tabular}[c]{@{}c@{}}Kernel \\ Ridge \end{tabular}  
    & \begin{tabular}[c]{@{}l@{}}alpha\\ gamma\\ kernel  \end{tabular}                                   
    & \begin{tabular}[c]{@{}c@{}}\num{1e-5} - 1\\ 0.01 - 150\\ rbf, linear \end{tabular}                           
    & \begin{tabular}[c]{@{}c@{}}8\\ 10\\ 2 \end{tabular}     
    & \begin{tabular}[c]{@{}c@{}}\num{1e-5}\\ 5\\ rbf  \end{tabular}                
    \\ \hline
    \begin{tabular}[c]{@{}c@{}}MLP \\ Regressor \end{tabular}    
    & \begin{tabular}[c]{@{}l@{}}activation\\ alpha\\ hidden layer sizes\\ learning rate\\ solver\end{tabular}              
    & \begin{tabular}[c]{@{}c@{}}logistic, relu, tanh\\ \num{1e-4} - \num{5e-3}\\ nodes: 50 - 120, 2-4 layers\\ constant, invscaling, adaptive\\ sgd, adam, lbfgs\end{tabular}   
    & \begin{tabular}[c]{@{}c@{}}3\\ 7\\ 4\\ 3\\ 3\end{tabular}  
    & \begin{tabular}[c]{@{}c@{}}tanh\\ \num{1e-3}\\$[50,100,200,100,50]$\\ adaptive\\ lbfgs\end{tabular}    
    \\ \hline
    \end{tabular}
\end{table*}

Table~\ref{tab:hyper_time} compares the computation times required for hyperparameter optimization across the models.  
Due to the curse of dimensionality, models with more hyperparameters generally require much longer tuning times.  
For instance, the Random Forest model involves evaluating \num{19008} hyperparameter combinations ($9{\times}4{\times}11{\times}4{\times}3{\times}4{=}\num{19008}$), as detailed in Table~\ref{tab:hyperparameters}.  
In contrast, the Decision Tree and MLP Regressor models involve fewer combinations-600 and 756, respectively.  
However, the MLP Regressor still requires significantly more tuning time due to the higher training cost per configuration.  
Unlike Decision Trees, MLPs must learn additional internal parameters, such as weights and biases, for each hyperparameter setting~\cite{goodfellow2016}. 
In total, hyperparameter tuning for the MLP Regressor takes over \SI{12}{\hour}, whereas for the Decision Tree model it takes only about \SI{16}{\second}, a negligible amount by comparison.

\begin{table*}[thbp]
    \centering
    \caption{Assessment of time required for performing hyperparameter optimizations.}
    \label{tab:hyper_time}
    \begin{tabular}{@{}cccccccccc@{}}
    \toprule
        & \multicolumn{1}{c}{\begin{tabular}[c]{@{}c@{}}Decision\\ Tree\end{tabular}} 
        & \multicolumn{1}{c}{\begin{tabular}[c]{@{}c@{}}Random\\ Forest\end{tabular}} 
        & AdaBoost 
        & \multicolumn{1}{c}{\begin{tabular}[c]{@{}c@{}}Gradient\\ Boost\end{tabular}}
        & XGBoost
        & \multicolumn{1}{c}{\begin{tabular}[c]{@{}c@{}}SVM\\ Regressor\end{tabular}}
        & \multicolumn{1}{c}{\begin{tabular}[c]{@{}c@{}}GP\\ Regressor\end{tabular}}
        & \multicolumn{1}{c}{\begin{tabular}[c]{@{}c@{}}Kernel\\ Ridge\end{tabular}}
        & \multicolumn{1}{c}{\begin{tabular}[c]{@{}c@{}}MLP\\ Regressor\end{tabular}}
        \\ \midrule
        \begin{tabular}[c]{@{}c@{}}Tuning \\ time (\SI{}{\second}) \end{tabular} 
        & \num{16}                                                                           
        & \SI{1628304}{}                                                  
        & \SI{42814}{} 
        & \SI{900946}{} 
        & \SI{120845}{} 
        & \SI{25120}{} 
        & \SI{61494}{} 
        & \SI{2599}{} 
        & \SI{44325}{} 
        \\ \bottomrule
    \end{tabular}
\end{table*}
\subsection{Accuracy Metrics}\label{sec_accuracy_metrics}
To evaluate the performance of the models, four metrics are introduced: normalized mean square error (nMSE), 
normalized mean absolute error (nMAE), normalized max-error (nMaxE), and the $R^2$ score~\cite{Botchkarev2019}. Accuracy metrics
without normalization are scale-dependent~\cite{hyndman2006}. In other words, scale-dependent measures always require 
information about the scale of the target to assess the model's accuracy effectively. For the following expressions of normalized metrics, 
$A_e$ denotes the ground truth, $\hat{A}_e$ is the prediction, $\bar{A}_e$ is the mean ground truth of the test data, 
$\max (|A_e|)$ is the max value of ground truth in the test data, and $n$ is the number of test points. 

The normalized mean square error (nMSE) is defined as: 
\begin{equation}
    \text{nMSE} = 100 \ \frac{\frac{1}{n} \sum\limits_{i=1}^n \bigl(A_e^i - \hat{A}_e^i \bigr)^2}{\bar{A}_e} \%.
\end{equation}
The nMSE metric quantifies the average squared differences between predictions and ground truths, normalized by the mean of the ground truth. 
Since it penalizes larger errors more heavily, it is particularly sensitive to outliers. A lower nMSE indicates a better model fit.

The normalized mean absolute error (nMAE) is formulated as: 
\begin{equation}
    \text{nMAE} = 100 \ \frac{\frac{1}{n} \sum\limits_{i=1}^n \big|A_e^i - \hat{A}_e^i\big|}{\bar{A}_e} \%.
\end{equation}
The nMAE metric measures the average absolute differences between predictions and ground truths, providing an interpretable error metric. 
Unlike nMSE, nMAE treats all errors equally and is more robust to outliers. Lower nMAE values indicate higher prediction accuracy.

The normalized max-error (nMaxE) is expressed as: 
\begin{equation}
    \text{nMaxE} = 100 \ \frac{\max \Bigl(\big|\boldsymbol{A_e} - \boldsymbol{\hat{A}_e}\big|\Bigr)}
                              {\max \Bigl(\big|\boldsymbol{A_e}\big|\Bigr)} \%.
\end{equation}
The nMaxE metric captures the largest absolute error relative to the maximum observed value in the test data. 
It highlights the worst-case error scenario, making it useful for applications where large deviations are critical. 
A lower nMaxE suggests that the model avoids large prediction errors.

The $R^2$ metric, also referred to as the coefficient of determination, quantifies how well the predictions match with actual values.
Usually, the $R^2$ value ranges from 0 to 1, or from 0\% to 100\%. The accuracy metric $R^2$ is defined as
\begin{equation}
    R^2 = 100 \Biggl[ 1 - \frac{\sum\limits_{i=1}^n \bigl(A_e^i - \hat{A}_e^i\bigr)^2}{\sum\limits_{i=1}^n \bigl(A_e^i - \bar{A_e}\bigr)^2} \Biggr] \%.
\end{equation}
$R^2$ measures the proportion of variance in the actual values that is explained by the model predictions. 
A value close to 100\% indicates a strong predictive model, while lower values suggest poor explanatory power.

Accuracy metrics are utilized for quantifying the investigated surrogate models based on the predicted and actual 
effective contact area. Also, note that all metrics are unitless and defined as percentages.  
\section{Results and Discussion}\label{sec_results}
\subsection{Evaluation of Baseline Models}\label{sec_baseline}
The evaluation of the so-called baseline models aims to compare the investigated machine learning algorithms without hyperparameter optimization
in terms of accuracy metrics, fit, and prediction time on unseen test data.     

\begin{figure*}[thbp]
    \centering
    \includegraphics[width=0.9\textwidth]{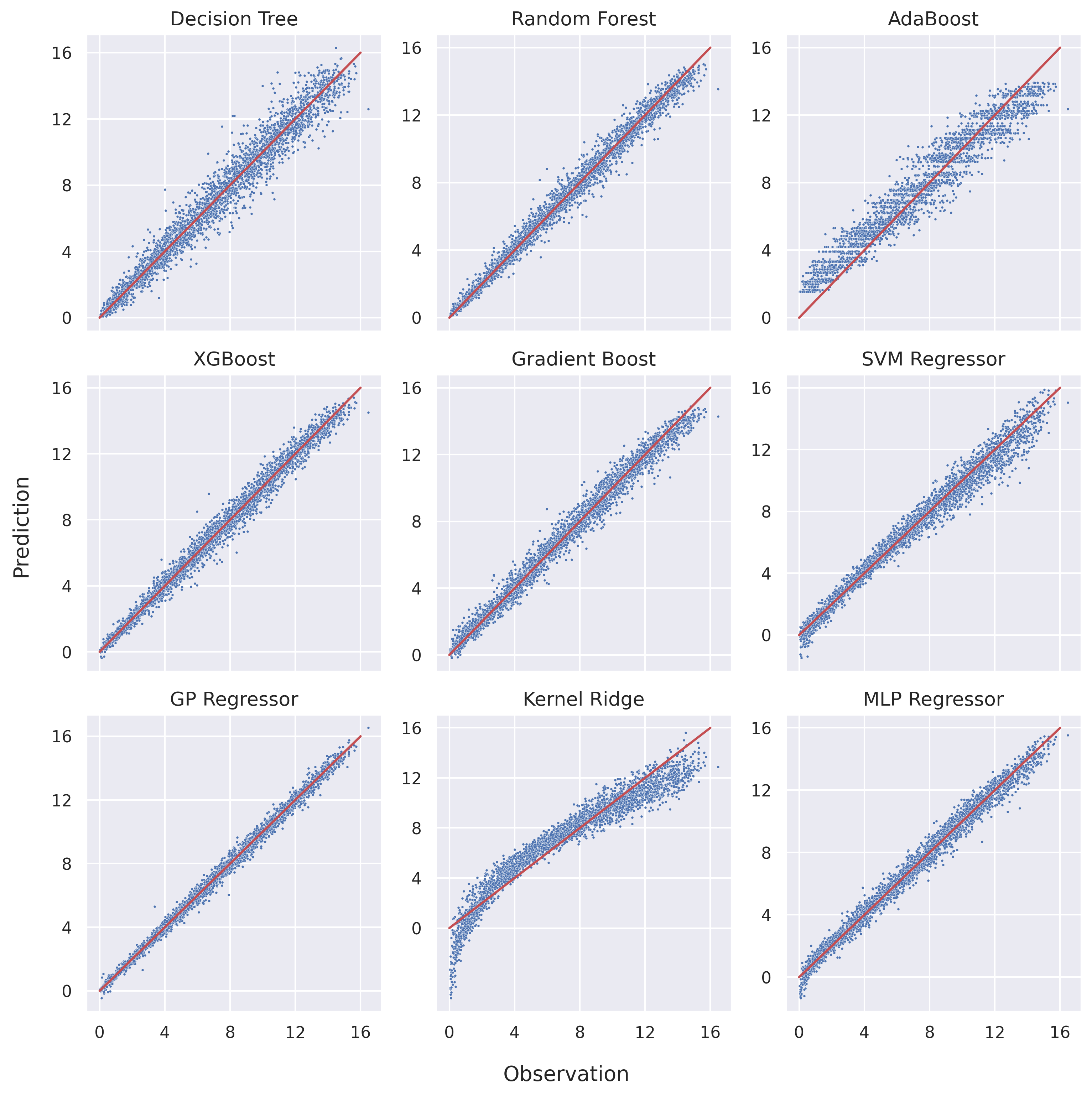}
    \caption{Comparison of the investigated baseline models to predict the effective contact area (\%). 
    Horizontal axes are the observations or actual values, while vertical axes denote the predictions. 
    The red lines represent a perfect fit, which is used to qualify the model's accuracy.}
    \label{fig:baseline_comparison}
\end{figure*}

Figure~\ref{fig:baseline_comparison} shows the qualitative comparison of the investigated baseline models to predict 
the effective contact area. Each figure has a red line, also referred to as the perfection line, 
which is used to qualify the model prediction. 
The qualitative accuracy of the model can be determined by how closely its predictions align with the red curve.
A simple tree-based method such as Decision Tree shows a linear trend that is consistent with the perfection line. However, 
it exhibits a high variance, which causes a bad generalization. On the other hand, ensemble methods that rely on multiple
decision trees, such as Random Forest, Gradient Boost, and XGBoost, exhibit reduced variance, thus leading to improved accuracy.   
While the majority of ensemble methods perform well, AdaBoost stands out as a negative exception.
The AdaBoost algorithm struggles to grasp the underlying patterns in the data, 
often delivering the same predictions although the actual values are different, thus 
indicating potential underfitting.

The SVM Regressor deviates slightly from the perfection line with an increasing variance for larger values of the effective contact area. 
In cases where the target values are small, the SVM Regressor tends to produce some negative, physically unrealistic predictions 
(the effective contact area can not take negative values). 
Similarly, the Kernel Ridge algorithm excessively deviate 
from the perfection line, predicting large negative values of the effective contact area 
even though the actual values are meant to be small yet positive. 
These noticeable deviations around small effective contact area values can be related to the dominating effect 
of the loss function contributions of data points with large effective contact area values during training. Since the Kernel Ridge algorithm is not yet tuned and has a simple 
model complexity, it tends to prioritize minimizing errors for larger values, which is why we observe 
significant deviations around the small effective contact values. 
On the other hand, the MLP Regressor with a single hidden layer using 
the rectified linear unit (relu) activation function demonstrates a behavior similar to that of the SVM Regressor,
In particular, it tends to predict comparably large negative values of the effective contact area around the origin. However,
in contrast to the SVM Regressor, exhibits significantly lower variance, especially for larger effective contact areas.
The GP Regressor model outperforms all other baseline models 
in terms of accuracy and exhibits comparably reduced variance around the perfection line.     

\begin{figure*}[thbp]
    \centering
    \stackinset{c}{1.2cm}{t}{-0.25cm}{(a) \hspace*{6cm} (b)}%
    {\includegraphics[width=0.95\textwidth]{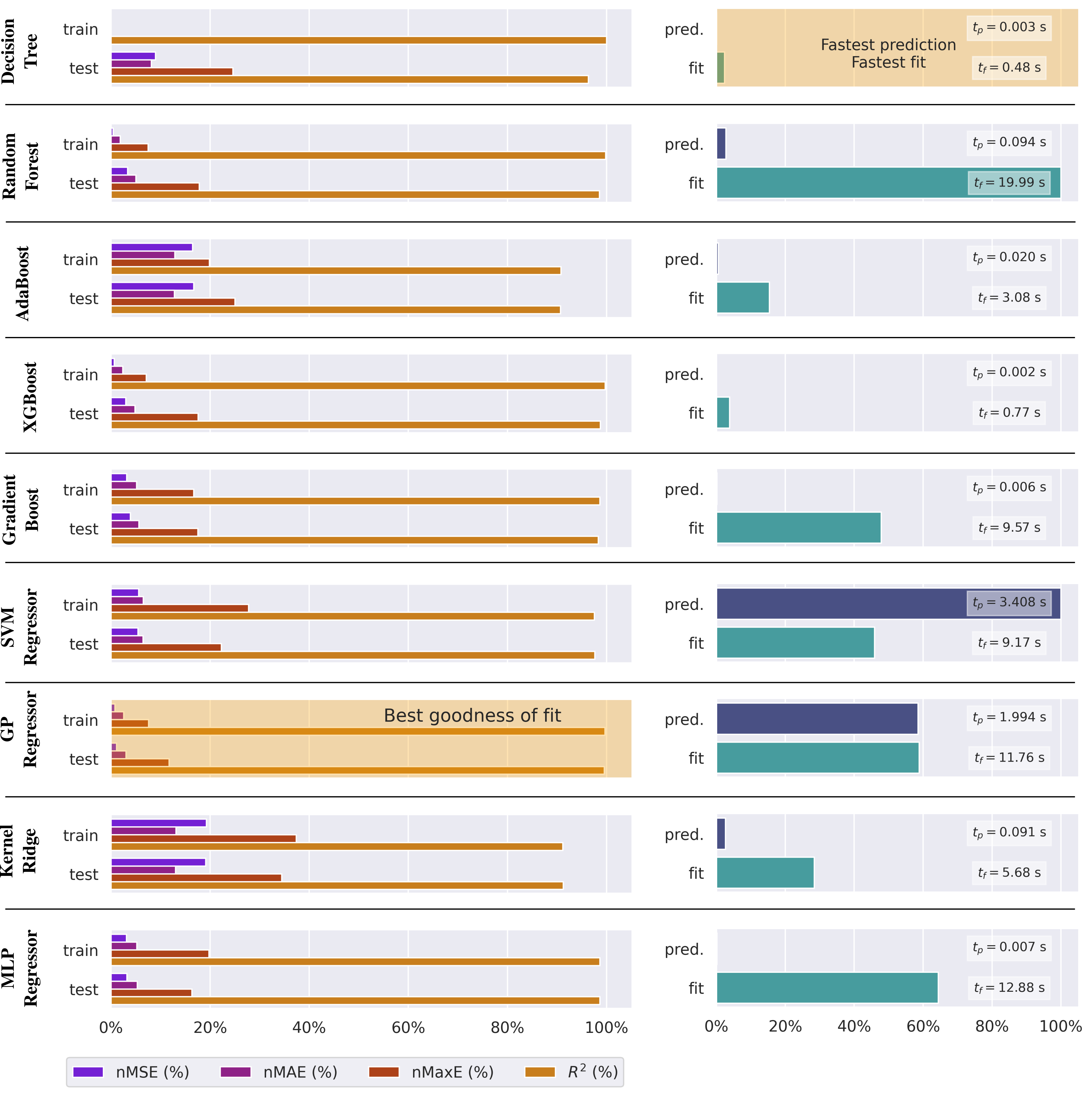}}
    \caption{Comparison of the investigated data-driven baseline models to predict the 
    effective contact area for rough surface contact: (a) Accuracy metrics (\%),
    (b) times for fit and prediction in \% are given including the corresponding actual values in s. 
    Note that each accuracy metric is scaled relative to the worst accuracy across all models to enable better comparisons.
    Similarly, each prediction and fit time measures are scaled relative to the maximum prediction and fit time across all models. 
    }
    \label{fig:baseline_metrics}
\end{figure*}

Figure~\ref{fig:baseline_metrics}a provides a quantitative comparison of accuracy metrics, 
namely nMSE (\%), nMAE (\%), nMaxE (\%), and the $R^2$ score (\%) for the investigated baseline models. Since the Decision Tree 
algorithm by definition mimics the training data, it yields 0\% nMSE, nMAE, nMaxE along with 100\% $R^2$ in the training data set. However, 
it suffers from a relatively high nMaxE value in the test data, e.g., \nMaxE{24.62}. The corresponding numbers for nMSE, nMAE, nMaxE and $R^2$ are provided in Table~\ref{tab:comparison}.
The large difference between train and test metrics means that the baseline model based on the Decision Tree 
can not properly generalize the problem. 
Yet, the Decision Tree model has the fastest prediction and fit times as depicted in Figure~\ref{fig:baseline_metrics}b. 
The Random Forest exhibits smaller test errors in terms of nMSE and nMAE compared to nMaxE and it does not indicate any sign of
overfitting. However, it has the largest fit time of \fitTime{19.99}. Similarly, the SVM Regressor exhibits larger nMaxE values 
when compared to other accuracy metrics and it has the largest prediction time with \predTime{3.408}. 
The Kernel Ridge model produces the largest test errors in the test data with \nMSE{19.11}, \nMAE{13.02}, and \nMaxE{34.45}. Moreover,
it has a poor performance in terms of $R^2$. The ensemble method AdaBoost also produces large test errors and has the poorest 
coefficient of determination of \RScore{90.71}. Other ensemble methods such as Gradient Boost and XGBoost demonstrate better results
in general. Additionally, XGBoost has the second-fastest fit and prediction time, following closely after the Decision Tree.
As a neural network model, MLP Regressor produces smaller test errors in terms of nMSE and nMAE, but it has the largest fit time with \fitTime{12.88}. 
Among all the investigated methods, the GP Regressor model stands out for its exceptional performance 
when considering various accuracy metrics in test data, e.g., \nMSE{1.15}, \nMAE{3.05}, and \nMaxE{11.76} along with 
\RScore{99.53}.
\subsection{Evaluation of Tuned Models}\label{sec_hyper}
\begin{figure*}[thbp]
    \centering
    \includegraphics[width=0.9\textwidth]{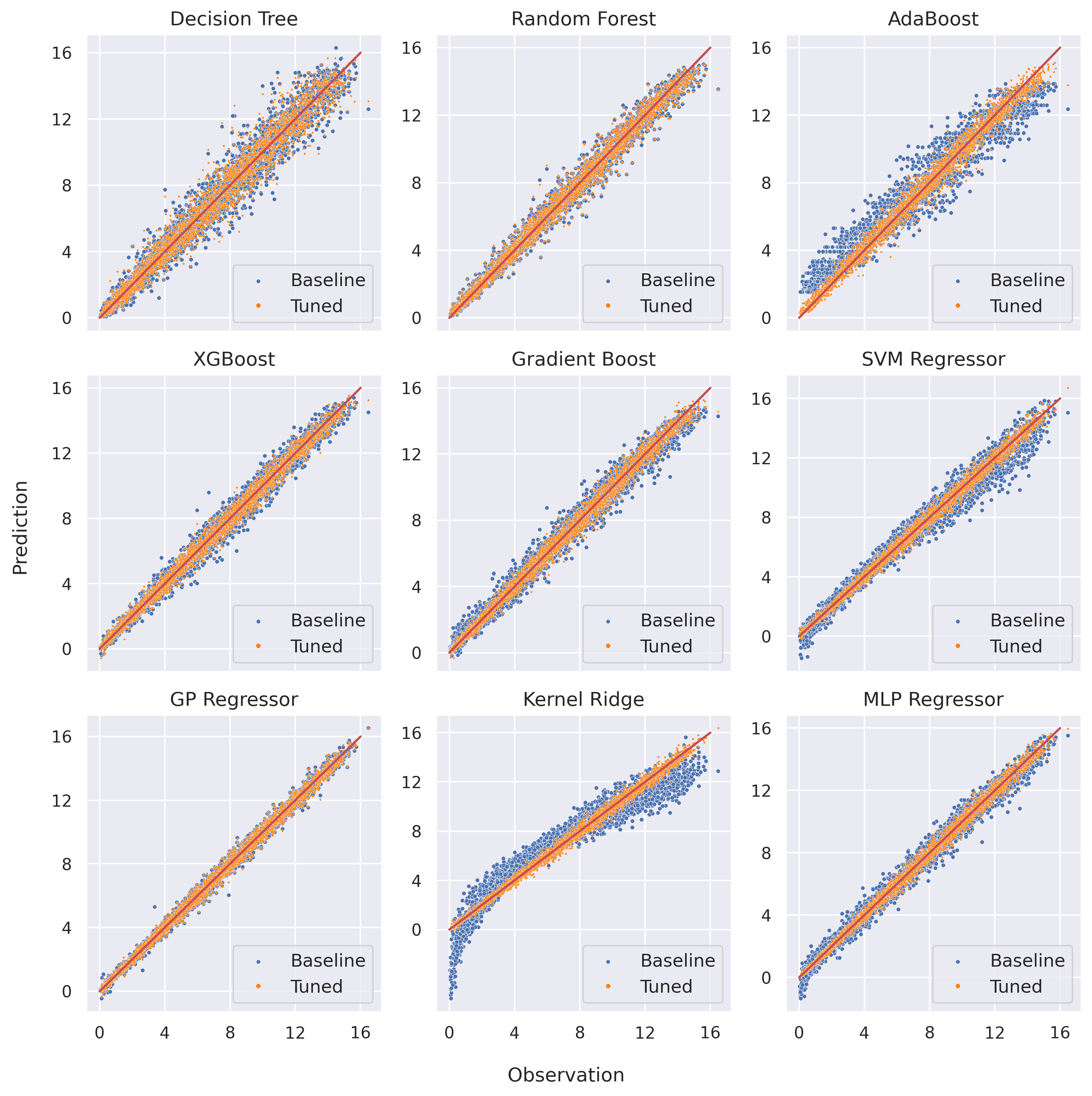}    
    \caption{Comparison of the baseline models with the tuned models. Horizontal axes represent the observations or actual values, 
    while vertical axes denote the predictions. The red lines represent a perfect fit, which is used to qualify
    the model's accuracy.}
    \label{fig:hyper_baseline_comparison}
\end{figure*}

As mentioned earlier, hyperparameter optimization or tuning is a crucial and necessary step to improve model performance 
and the ability to generalize from seen to unseen data. Also, tuning can reduce prediction and fit times. 
Figure~\ref{fig:hyper_baseline_comparison} qualitatively illustrates the comparison of baseline models and tuned models with their respective predictions of the effective
contact area. For all investigated models, there is a drastic improvement in accuracy. Particularly for 
AdaBoost, SVM Regressor, Kernel Ridge, and MLP Regressor models, predictions now closely align with the perfection line
in contrast to the baseline models. 
Furthermore, the tuned models are quite effective in avoiding unphysical negative values for the effective contact area,
particularly for the Kernel Ridge, the SVM Regressor and the MLP Regressor.

Table~\ref{tab:comparison} provides quantitative comparisons of baseline and tuned models in terms of 
accuracy metrics and computational time for model fitting and prediction. 
Hyperparameter optimization yields varying improvements across models, often at the expense of increased training time.
For both the Decision Tree and Random Forest models, tuning yields slight improvements in most accuracy metrics, except 
for nMaxE. Specifically, for the Decision Tree, the prediction time remains almost the same, while the fit time 
decreases slightly from \SI{0.48}{\second} to \SI{0.42}{\second}.
Conversely, the Random Forest model shows a drastic increase in both training and prediction time, 
nearly 15 times larger. 
The fit time increases dramatically from \SI{19.99}{\second} to \SI{298.58}{\second} and the prediction time 
rises from \SI{0.094}{\second} to \SI{1.364}{\second}. 
The tuned SVM Regressor demonstrates remarkable performance improvements, with nMSE decreasing from
from 5.47\% to 1.20\%, nMAE from 6.50\% to 3.09\%, and nMaxE from 22.30\% to 10.62\%, while the $R^2$ value increases from 
97.61\% to 99.50\%. However, these gains come with a steep computational cost, as the fit time increases from 
\SI{9.17}{\second} to \SI{2550}{\second}.
Yet, the tuned SVM Regressor model exhibits a reasonable reduction in prediction time. 
Tuned ensemble methods, e.g., AdaBoost, Gradient Boost, 
and XGBoost, exhibit great improvements in accuracy metrics with an average improvement factor of approximately $40$\%.
Among the ensemble approaches, the tuned XGBoost model demonstrates outstanding computing performance, delivering a fit time of \fitTime{4.13} and 
a prediction time of \predTime{0.005}.

The tuned MLP Regressor model exhibits better goodness of fit than the tree-based models across all accuracy metrics, with \nMSE{1.19}, \nMAE{3.09}, \nMaxE{9.95}, and \RScore{99.51}.  
With these values, it ranks as the third-best model overall in terms of accuracy.  
However, training the tuned MLP Regressor requires approximately \SITime{3435}, making it the most time-consuming training process  among all models.
The fit time is approximately 266 times longer than that of the baseline model, primarily due to a significantly higher number of training epochs and increased model complexity.  
Specifically, the tuned model is trained over \SI{20000} epochs, compared to just 200 epochs for the baseline.  
In addition, the tuned MLP employs a deeper architecture with more neurons and layers, resulting in a substantially larger number of internal parameters to optimize.  
With a prediction time of \SI{0.049}{\second}, the MLP Regressor lies in the middle range.  
It is an order of magnitude slower than the Decision Tree, yet two orders of magnitude faster than the slowest models, such as the GP Regressor and Random Forest.  

For the GP Regressor model, tuning leads to slight improvements in nMSE and nMAE, and a substantial reduction in nMaxE---from 11.76\% to 8.98\%.  
As a result, the GP Regressor achieves the best overall accuracy across all evaluated metrics, making it the most effectively tuned model in this study.  
However, this gain in accuracy comes at a considerable computational cost: the fit time increases sharply from \SI{11.76}{\second} to \SI{853.23}{\second}, representing more than a 70-fold increase, while the prediction time remains nearly unchanged.  
A significant portion of this fit time is spent on computing the maximum likelihood estimation of the kernel parameters.  
In contrast, the tuned Kernel Ridge model delivers accuracy metrics comparable to those of the GP Regressor, but with much lower computational demands-requiring only \SI{9.18}{\second} for training and \SI{0.827}{\second} for prediction.
Overall, we consider the tuned Kernel Ridge model to be the most balanced surrogate, as discussed in Section~\ref{sec_discussion}.  
For this reason, we focus on this model in greater detail in the following sections.

\begin{table*}[thbp]
    \definecolor{taborange}{HTML}{FF7F0E}
    \definecolor{tabblue}{HTML}{1f77b4}
    \definecolor{tabgreen}{HTML}{2ca02c}
    \def\cellcolorbase{\cellcolor{tabblue!75}}
    \def\cellcolortune{\cellcolor{taborange!75}}
    \def\cellcoloroptimal{\cellcolor{tabgreen!75}}
    \centering
    \caption{Comparison of the baseline models with the tuned models. Accuracy metrics nMSE, nMAE, nMaxE, $R^2$ are evaluated on the test data.
    Improvement is calculated based on $100(m_t - m_b)/m_b$ where $m_b$ is the evaluated performance metric of the baseline model 
    and $m_t$ is the evaluated performance metric of the tuned model.
    The cells with {\color{tabblue}blue} background show the model accuracy of the best baseline model, 
    the cells with {\color{taborange}orange} background denote those of the best-tuned model, 
    while the cells with {\color{tabgreen}green} background show the model accuracy and performance of the model considered "optimal" overall.
    The model accuracy is determined based on accuracy metrics, and the model performance is assessed by examining fit and prediction time. 
    The fit (or training) time refers to the time required to train the models on the training dataset, whereas the prediction time is measured on the test dataset and represents the total time needed to make predictions for all test data points.}
    \label{tab:comparison}
    \begin{tabular}{@{}cl@{\hskip 0.75cm}llll@{\hskip 1cm}ll@{}}
    \multicolumn{1}{l}{}                                                     &             & \multicolumn{4}{c}{\textbf{Performance metric (\%)}} & \multicolumn{2}{c}{\textbf{Time (s)}} \\ \cmidrule(r{1cm}){3-6} \cmidrule{7-8}
    \multicolumn{1}{l}{}
                                                                             &             & nMSE    & nMAE    & nMaxE    & $R^2$   & fit ($t_f$)            & pred. ($t_p$)       \\ \midrule
    \textbf{%
    \multirow{3}{*}{\begin{tabular}[c]{@{}c@{}}Decision\\ Tree\end{tabular}}}& Baseline    & 8.99    & 8.17    & 24.62    & 96.31&  0.48           &  0.003            \\
                                                                             & Tuned       & 8.51    & 8.03    & 27.66    & 96.51&  0.42           &  0.003            \\
                                                                             & Improvement & 5.34    & 1.71    & -12.35    & 0.21 &              &            \\ 
    \midrule    
    \textbf{%
    \multirow{3}{*}{\begin{tabular}[c]{@{}c@{}}Random\\ Forest\end{tabular}}}& Baseline    & 3.39    & 5.05    & 17.81    & 98.56& 19.99          & 0.094            \\
                                                                             & Tuned       & 3.34    & 5.01    & 18.58    & 98.58& 298.58         & 1.364           \\
                                                                             & Improvement & 1.47    & 0.79    & -4.32   &  0.02 &                &             \\ 
    \midrule
    \textbf{%
    \multirow{3}{*}{AdaBoost}}                                               & Baseline    & 16.71   & 12.80   & 25.03    & 90.71 & 3.08           & 0.020          \\
                                                                             & Tuned       & 2.92    & 4.77    &  16.34   & 98.76 & 230.11        & 1.174            \\
                                                                             & Improvement & 82.53   & 62.73   &  37.72   &  8.87 &                &             \\ 
    \midrule
    \textbf{%
    \multirow{3}{*}{XGBoost}}                                                & Baseline    & 3.01    & 4.86    & 17.59    & 98.75 & 0.77           & 0.002            \\
                                                                             & Tuned       & 1.95    & 4.03    & 10.34    & 99.19 & 4.13           & 0.005            \\
                                                                             & Improvement & 35.22   & 17.08   & 41.25    &  0.45 &                &             \\ 
    \midrule
    \textbf{%
    \multirow{3}{*}{\begin{tabular}[c]{@{}c@{}}Gradient\\Boost\end{tabular}}}& Baseline    & 3.95    & 5.68    & 17.55    & 98.30 & 9.57           & 0.006          \\
                                                                             & Tuned       & 2.03    & 4.03    & 11.55    & 99.15 & 197.10         & 0.067          \\
                                                                             & Improvement & 48.61   & 29.05   & 34.19    &  0.86 &                &             \\ 
    \midrule
    \textbf{%
    \multirow{3}{*}{\begin{tabular}[c]{@{}c@{}}SVM\\ Regressor\end{tabular}}}& Baseline    & 5.47    & 6.50    & 22.30    & 97.61 & 9.17           & 3.408            \\
                                                                             & Tuned       & 1.20    & 3.09    & 10.62    & 99.50 & 2550.27       & 0.624           \\
                                                                             & Improvement & 78.06   & 52.46   & 52.38    &  1.94 &                &             \\ 
    \midrule
    \textbf{%
    \multirow{3}{*}{\begin{tabular}[c]{@{}c@{}}GP\\ Regressor\end{tabular}}} & Baseline    & \cellcolorbase 1.15 & \cellcolorbase 3.05 & \cellcolorbase 11.76 & \cellcolorbase 99.52 & 11.76 & 1.994        \\
                                                                             & Tuned       & \cellcolortune 1.12 & \cellcolortune 3.02 & \cellcolortune  8.98 & \cellcolortune 99.54 & 853.23  & 1.965      \\   
                                                                             & Improvement & 2.61                & 0.98                &  23.64               &  0.02                &         &            \\ 
    \midrule
    \textbf{%
    \multirow{3}{*}{\begin{tabular}[c]{@{}c@{}}Kernel\\ Ridge\end{tabular}}} & Baseline    & 19.11   & 13.02   & 34.45    & 91.24 & 5.68           & 0.091            \\
                                                                             & Tuned       & \cellcoloroptimal 1.16    &\cellcoloroptimal 3.07    &\cellcoloroptimal 9.84     &\cellcoloroptimal 99.52&\cellcoloroptimal 9.18           &\cellcoloroptimal 0.827           \\
                                                                             & Improvement & 93.93   & 76.44   & 71.44    & 9.07 &                &             \\ 
    \midrule
    \textbf{%
    \multirow{3}{*}{\begin{tabular}[c]{@{}c@{}}MLP\\ Regressor\end{tabular}}}& Baseline    & 3.24   & 5.32    & 16.34    & 98.64 & 12.88          & 0.007          \\
                                                                             & Tuned       & 1.19    & 3.09    & 9.95     & 99.51& 3435.54        & 0.049            \\
                                                                             & Improvement & 63.27   & 41.92   & 39.11    & 0.88 &                &             \\ 
    \midrule
   
    \end{tabular}
\end{table*}
\subsection{Evaluation of Kernel Ridge on a Fine Grid}
\begin{figure*}[thbp]
    \begin{minipage}{.7\textwidth}
    \centering
    \includegraphics[width=1\textwidth]{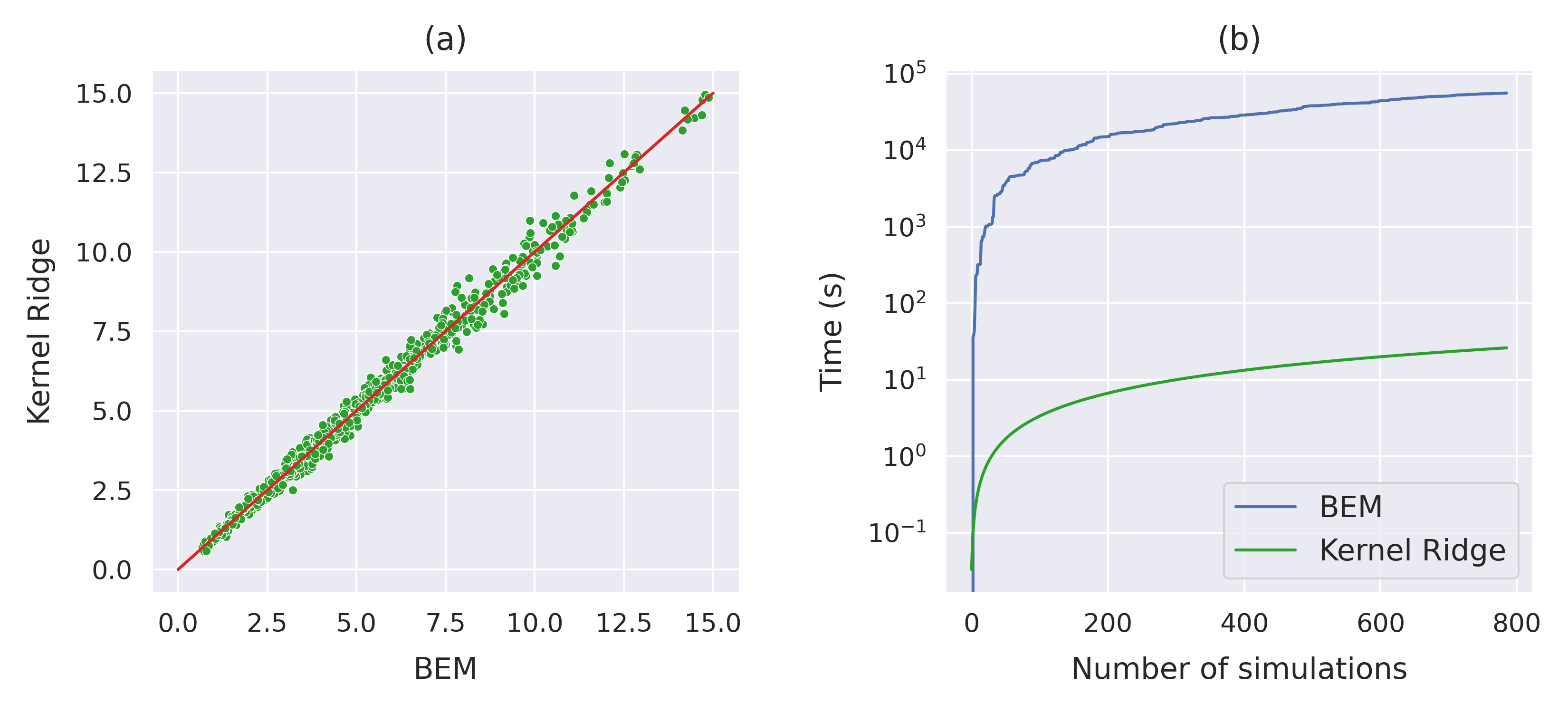} 
    \captionof{figure}{Comparison of the Kernel Ridge and the BEM model in terms of (a)
    effective contact area, (b) evaluation or prediction time. Rough surfaces are characterized by a $256 \times 256$ grid.}
    \label{fig:bem_kr_n_8}
    \end{minipage}\hspace{0.3cm}
    \begin{minipage}{.25\textwidth}
        \centering 
        \captionof{table}{Corresponding accuracy metrics.}
        \begin{tabular}[b]{@{}ll@{}}
                    \textbf{Metric (\%)} & \textbf{Value} \\ \midrule
                    nMSE   & 1.46      \\
                    nMAE   & 3.95      \\
                    nMaxE  & 7.56      \\
                    $R^2$  & 99.16     \\ \bottomrule
        \end{tabular}%
    \label{tab:bem_kr_n_8}
    \end{minipage}
\end{figure*}

As explained in Section~\ref{sec_data_generation}, the database is generated with rough contact simulations 
on a $128 \times 128$ grid. Correspondingly, surrogate models are trained using the statistical features 
that are obtained from these relatively coarse grids. The question naturally arising here is whether it would be possible to predict 
the effective contact area resulting from a finer grid, e.g., $256 \times 256$ points per side, 
using the surrogate Kernel Ridge model, which has been trained 
on a database whose features are extracted exclusively from samples having a coarser grid, namely $128 \times 128$. 
To answer this question, 800 additional BEM simulations based on a $256 \times 256$ grid are run. Following the same process as 
explained in Section~\ref{sec_data_generation}, a database for testing is generated including model parameters and statistical parameters based on the fine grids, 
and predictions are computed using the already trained Kernel Ridge model.
As depicted in Figure~\ref{fig:bem_kr_n_8}a, 
the predicted effective contact area via Kernel Ridge aligns nicely with the actual values obtained through
the BEM model, which is also confirmed by the computed accuracy metrics as provided in Table~\ref{tab:bem_kr_n_8}.
Indeed, compared to the coarser BEM grids, the computed accuracy metrics are slightly different, since no information from finer grids
is used during the model training. In terms of evaluation or prediction time, the surrogate model is more efficient 
and the gap between Kernel Ridge and BEM increases drastically since simulations with the finer grid take extensively 
longer computing time. 

\subsection{Comparative Cost Analysis of Reference and Surrogate Models}\label{sec_comparative_cost}
The motivation behind building surrogate models is to construct fast-to-evaluate approximations of the BEM reference model, thereby making multi-query scenarios, requiring numerous model evaluations, computationally tractable.  
While surrogate models reduce evaluation time compared to the reference model, they also introduce additional upfront costs associated with model construction.  

Ultimately, our goal is to determine the break-even point, the number of model evaluations at which the computational savings from using the surrogate outweigh the cost of building it.  
To analyze this, we conduct a comparative cost analysis of the overall computational cost of using the Kernel Ridge model versus the reference numerical method (BEM) for an increasing number of simulations.  
For the BEM model, the total cost $t_{\text{BEM}}$ is obtained by summing the time required for each individual simulation.  
As shown in Figure~\ref{fig:simulation_time}, the simulation time depends on both surface properties and the applied far-field displacement.  
Therefore, the exact cost of a specific BEM simulation is not known in advance. 
To approximate the total cost, we use the average simulation time from the database generation process, which is \SI{189,79}{\second}.  
This average is assumed to be a reliable estimate of expected cost, especially when considering a large number of simulations.

In contrast, computing the total cost for the surrogate model is more intricate.  
To ensure a fair comparison, all components of the surrogate modeling process must be considered.  
Specifically, we account for four time contributions involved in building and using the surrogate model.
The first contribution is the model evaluation time, which refers to the total cost $t_p$ of making predictions using the surrogate model across all input samples.  
To accurately estimate the prediction time for a single sample, we compute the average prediction time based on 2,000 randomly selected input samples, drawn uniformly from the feature ranges specified in Table~\ref{tab:list_of_params}.  
The second contribution is the fit time $t_f$, i.e., the time required to train the surrogate model on the available training data.  
The third contribution is the cost of hyperparameter optimization $t_h$, which, as shown in Table~\ref{tab:hyper_time}, can be substantial.  
Finally, we include the cost of generating the database $t_d$, as this is a necessary prerequisite for training any surrogate model. 
This includes both the generation of input samples (e.g., synthetic surfaces and far-field displacements) and the corresponding evaluations using the reference BEM model to obtain the output quantities.  
The total cost of the Kernel Ridge Regressor can finally be computed as $t_{\text{KRR}} = t_p + t_f + t_h + t_d$.

\begin{figure}[thbp]
    \def\hdistfig{4.6cm}
    \centering
    {\includegraphics[width=\columnwidth]{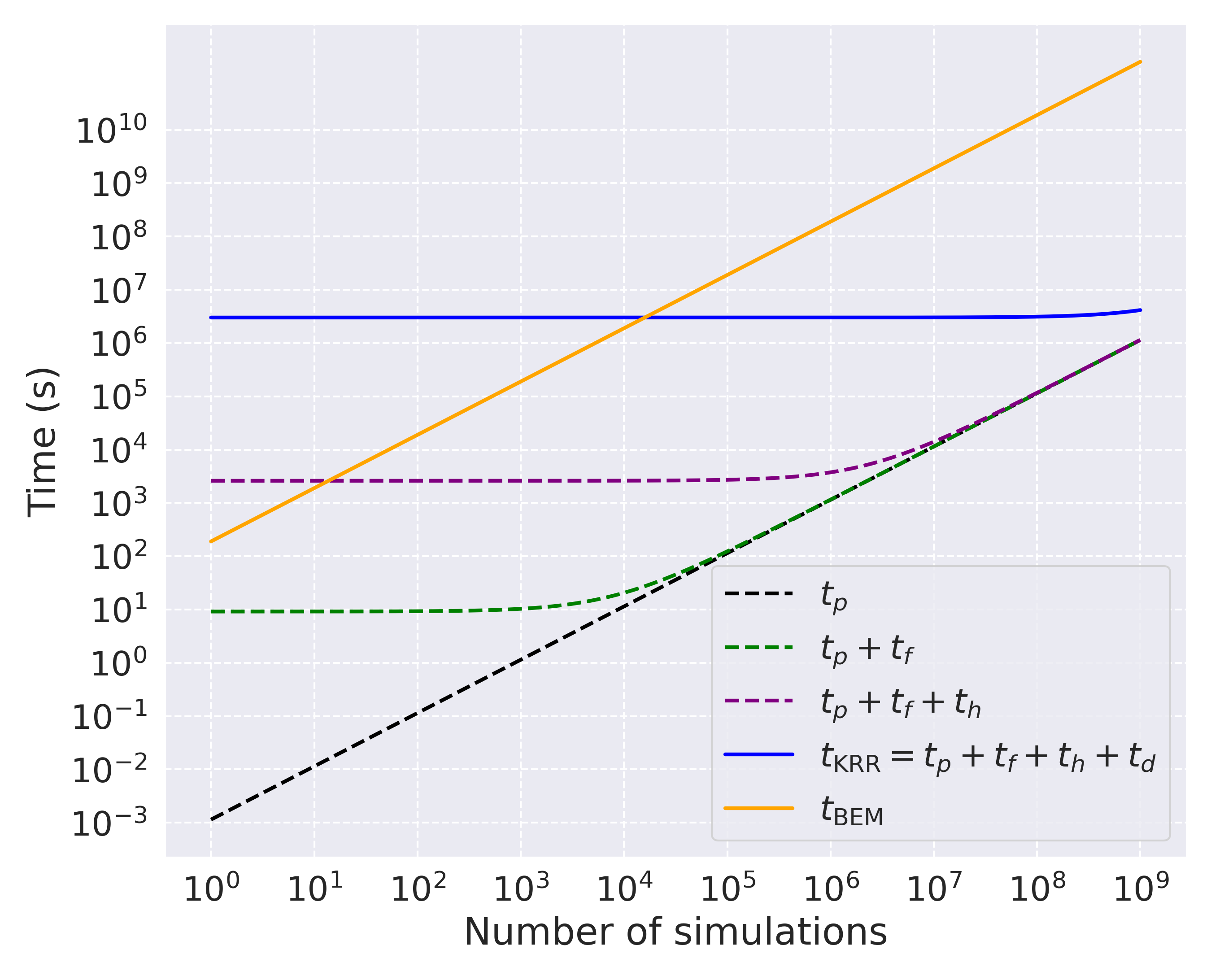}}
    \caption{Comparison of the total computational cost for the BEM model and the Kernel Ridge surrogate model.
    For BEM, the cost is estimated using the average simulation time (\SI{189.79}{\second}) per sample.
    The surrogate model cost includes four components: prediction time ($t_p$), training time ($t_f$), hyperparameter optimization time ($t_h$), and database generation time ($t_d$), with total cost $t_{\text{KRR}} = t_p + t_f + t_h + t_d$. 
    The surrogate model's overall cost is dominated by database generation ($t_d$), while training and hyperparameter optimization contribute marginally.
    The surrogate becomes more cost-efficient than BEM after approximately 15,893 simulations.
    }
    \label{fig:compare_bem_kernelridge}
\end{figure}

A visualization of the comparative computational cost analysis is provided in Figure~\ref{fig:compare_bem_kernelridge}.  
The overall cost of using the surrogate model is dominated by the database generation time $t_d$, which is several orders of magnitude greater than the other cost components.  
The model training time is negligible: on average, a single BEM evaluation is more expensive than the combined cost of training and evaluating the Kernel Ridge model.  
Even the hyperparameter optimization step, while computationally intensive within the surrogate modeling workflow, contributes only marginally when compared to the cost of evaluating the reference model.  
Its cost breaks even after just 14 model evaluations, including both training and prediction.  
A comparison of the total costs, $t_{\text{BEM}}$ and $t_{\text{KRR}}$, shows that the break-even point is reached at approximately 15,893 simulations.  
In other words, the surrogate model becomes the more cost-efficient choice only when more than this number of simulations is required.  
It is noteworthy that the database used to train the surrogate model consists of 15,878 samples, further underscoring the dominant cost associated with generating training data.  
Essentially, all other surrogate-related costs, including evaluation, training, and tuning times, are negligible in comparison.  
Although the break-even may initially seem like a high threshold, it is important to consider that applications such as uncertainty quantification (UQ) and Bayesian parameter calibration often involve several tens of thousands of model evaluations~\cite{Brandstaeter2021a}.  
In such scenarios, the break-even point can be reached relatively quickly, making surrogate modeling a practical and cost-effective solution.
\subsection{Discussion}\label{sec_discussion}
In Section~\ref{sec_baseline} and Section~\ref{sec_hyper}, the investigated surrogate models were compared in terms of accuracy-using standard metrics-and in terms of performance, based on model fit and prediction times.  
Our findings suggest that selecting an optimal surrogate model requires balancing accuracy with prediction performance, depending on the specific requirements of the application task.  
While the optimal choice is task-dependent, several general trends can be observed.  

Tree-based models, such as Decision Trees and Random Forests, typically perform worse in terms of accuracy compared to kernel-based methods.  
However, they offer significantly faster prediction times.  
This makes them suitable candidates in scenarios where rapid evaluations are critical and minor accuracy trade-offs are acceptable-such as in real-time applications.  
In contrast, kernel-based models achieve superior accuracy.  
Although their prediction times are generally higher than tree-based models, they are still relatively low in absolute terms (on the order of seconds for 3,175 test points), making them viable for most practical applications.  
Among the kernel-based models, the GP Regressor yields the best accuracy metrics, closely followed by the Kernel Ridge model.  
The differences in accuracy between these two models are negligible for most practical purposes.  
However, the GP Regressor has approximately twice the prediction time of the Kernel Ridge model.  
This is expected, as the GP Regressor also provides a predictive variance alongside the mean estimate---an added value that explains the increased computational cost.  
If uncertainty quantification is not required, the Kernel Ridge model provides nearly identical predictions at half of the computational cost.  
This is not surprising, as both methods are closely related.  
In fact, the mean prediction of the GP Regressor is mathematically equivalent to that of the Kernel Ridge model, provided the same hyperparameters are used---namely, the kernel type, kernel parameters, and regularization parameter.
The MLP Regressor, while slightly less accurate than the Kernel Ridge model, achieves lower prediction times---around one order of magnitude faster.  
However, this comes at the expense of a significantly longer training time and increased model complexity.  

Overall, the tuned Kernel Ridge model offers the best trade-off between accuracy and computational efficiency.  
It delivers high prediction accuracy with low prediction times and minimal training overhead, making it a strong candidate for general-purpose surrogate modeling.  
Therefore, while the GP Regressor achieves the highest accuracy, it is not considered the optimal model due to its substantially longer fit time and higher prediction cost.  
The tuned Kernel Ridge model is identified as the most balanced surrogate for a wide range of application scenarios.  

In Section~\ref{sec_comparative_cost}, we compared the overall costs of using a surrogate model with those of the reference model.  
Our results highlight that a fair comparison between data-driven surrogate models and reference models can only be made by considering the total cost---including both offline and online phases.  
The reasoning behind this is as follows:
A general machine learning regression task aims to learn an unknown functional relationship from input-output data.  
It is common practice to distinguish between the offline and online phases of model development, as they typically involve different tasks and cost structures~\cite{geron2022hands}. 
The offline phase includes data generation (collection, preprocessing, and feature selection) and model training.  
The online phase refers to the cost of evaluating the trained model.  
This separation is particularly relevant because online applications often involve a large number of model evaluations, while offline processes are performed only once.  
As a result, the substantial cost of the offline phase is usually justified as a one-time investment to obtain a usable model in the first place.

However, unlike standard machine learning scenarios where no prior model exists, surrogate modeling assumes the existence of a reliable reference model-typically one that is accurate but computationally expensive.  
The surrogate’s purpose is to approximate this reference model efficiently, enabling its use in applications such as optimization, uncertainty quantification (UQ), or parameter identification, where many evaluations are required.  
In these contexts, the total cost of using a surrogate model-including both offline and online phases-must be considered to fairly assess its efficiency in solving the application task.  
Solving the task with the reference model involves direct evaluation across many configurations, often resulting in high computational cost.  
In contrast, the surrogate model approach requires an initial investment: building the surrogate entails data generation, model training, and hyperparameter tuning.  
Once constructed, the surrogate is then evaluated across the same configurations as the reference model, but at significantly lower cost per evaluation.  
Ultimately, the initial cost of building a surrogate is incurred specifically for the application task at hand.  
Therefore, the offline cost must be included in the overall assessment of the surrogate model’s efficiency.  
Our results demonstrate that ignoring the offline costs leads to a distorted assessment, which falsely overstates the advantages of data-driven modeling.  
Comparing only the evaluation times of the models, as it is often done in the literature, is insufficient, since it neglects the largest cost component: database generation.  
However, the encouraging finding is that there is no practical need to rely on such an incomplete or unfair comparison.  
Even though the break-even point may initially appear high, our analysis shows that the proposed surrogate modeling approach remains highly worthwhile for many application tasks that involve large numbers of model evaluations.
\section{Conclusion} \label{sec_conclusion}
In this study, we developed a surrogate modeling framework to predict the effective contact area in rough surface contact problems involving a linear elastic half-space and a rigid rough surface.
A comprehensive database was constructed using BEM simulations, containing far-field displacements, 22 statistical parameters derived from synthetic surface height fields, and the target output-the effective contact area.
Several machine learning regression models were evaluated as surrogate modeling techniques.
Optimal model configurations were determined through hyperparameter optimization using grid search with cross-validation.
All models were assessed using standard accuracy metrics, as well as training and prediction times to evaluate their computational performance.

Our findings indicate that selecting an optimal surrogate model requires balancing predictive accuracy and computational performance, depending on the specific requirements of the application task.
Each model exhibits distinct characteristics, meaning that the ideal choice is context-dependent-for instance, a model prioritizing speed may be preferred in time-critical applications, whereas a more accurate model might be favored in precision-sensitive tasks.
Overall, our results show that the Kernel Ridge model offers the most favorable trade-off between accuracy and efficiency across a broad range of scenarios.
It achieves high accuracy, low prediction time, and minimal training overhead, making it a strong candidate for general-purpose surrogate modeling.
The Gaussian Process Regressor presents a compelling alternative when uncertainty quantification is required, although it incurs a higher computational cost due to variance estimation.
To assess the robustness of the surrogate modeling approach, the generalization capability of the Kernel Ridge was evaluated on previously unseen simulation scenarios.
The results demonstrate that the surrogate model maintains reliable predictive accuracy and computational efficiency even when applied beyond the resolution used during training, highlighting its potential for broader applicability.
When evaluating the effectiveness of the surrogate modeling approach, the break-even point---where the overhead of building the surrogate is offset by its reduced evaluation cost---is of particular interest.
We emphasize the importance of accounting for all offline costs, especially database generation.
Neglecting these costs, as is sometimes done in the literature, results in a skewed evaluation that overstates the benefits of data-driven modeling.
Nonetheless, we conclude that such an incomplete assessment is unnecessary in practice.
Even though the break-even point may initially appear high, the proposed surrogate modeling framework proves to be highly beneficial for many application scenarios that involve large numbers of model evaluations.

To conclude, the identified Kernel Ridge surrogate model predicts the effective contact area with acceptable accuracy and robustness while significantly reducing evaluation time compared to state-of-the-art numerical methods such as the Boundary Element Method (BEM).
As such, it serves as an attractive and efficient surrogate for use in multi-query analyses, such as uncertainty quantification, parameter identification, and sensitivity analysis.

This work reveals several opportunities for further investigations. Apart from statistical parameters, only 
the far-field displacement has been considered as input for the surrogate modeling strategy. 
However, the input space could be extended to other properties, e.g., geometrical  parameters such as the scan length.
Moreover, not only the effective contact area but also the overall contact force could be considered 
as relevant model output to formulate and solve the problem as a multi-target regression task. 
As an alternative approach, the contact force field can be predicted using a neural network approach consisting
of both fully connected and convolutional neural networks, by taking contour plots of the topographies as image-based input and 
far-field displacement values as numerical input. 
Eventually, the proposed surrogate modeling framework could be formulated as a hybrid model that combines data-driven components and model-based approaches that include the physics of the underlying problem.

\subsection*{Declarations}
\small
\textbf{Availability of data and materials}\\
The datasets generated and analysed during the current study, as well as the source code, 
are publicly available at \url{https://github.com/imcs-compsim/deep_sim}.
\vspace{0.2cm}

\noindent \textbf{Competing interests}\\
The authors declare that they have no competing interests.
\vspace{0.3cm}

\noindent \textbf{Funding}\\
\noindent This research paper is funded by dtec.bw - Digitalization and Technology Research Center of the 
Bundeswehr (project RISK.twin). dtec.bw is funded by the European Union - NextGenerationEU.
\vspace{0.3cm}


\noindent \textbf{Acknowledgments}\\
\noindent The authors gratefully acknowledge the computing resources provided by the Data Science \& 
Computing Lab at the University of the Bundeswehr Munich.

\begin{appendices}
\section{Comparison of GP Regressor and Kernel Ridge}\label{Appendix1}
\begin{figure*}[htbp]
    \centering
    \begin{minipage}[t]{\textwidth}
        \centering
        \includegraphics[width=\textwidth]{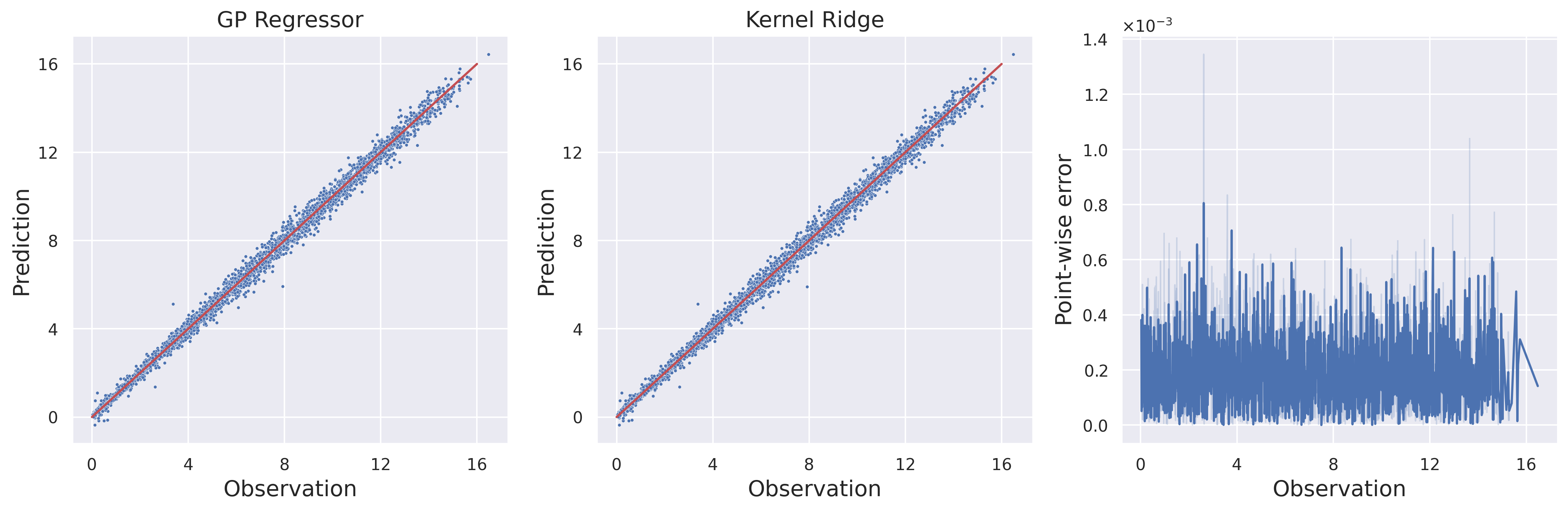}  
        \caption{Case 1: Good generalization.}
        \label{fig:figure1}
    \end{minipage}
    
    \vspace{0.3cm}  
    
    \begin{minipage}[b]{\textwidth}  
        \centering
        \includegraphics[width=\textwidth]{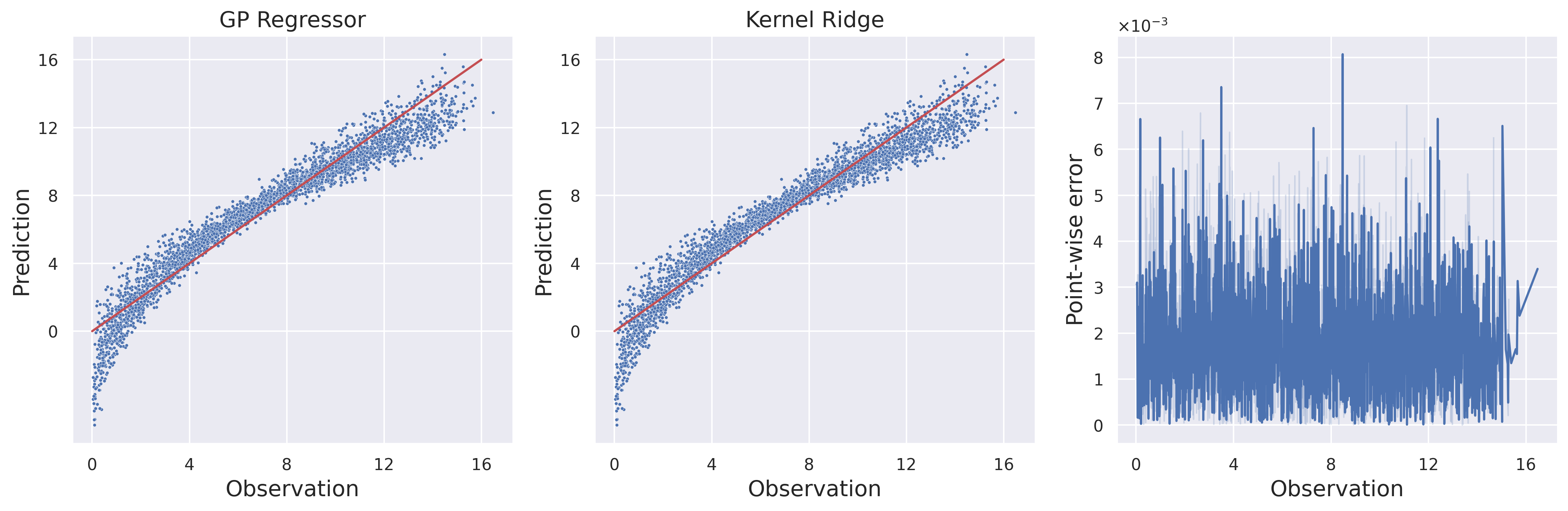}  
        \caption{Case 2: Bad generalization.}
        \label{fig:figure2}
    \end{minipage}
    \caption{Comparison of GP Regressor and Kernel Ridge under the assumption of both models having the same kernel
    including identical kernel parameters and the regularization parameter.}  
    \label{fig:compare_gp_kernelridge}
\end{figure*}

In the kernel-based regression, the predicted output $\hat y$ is calculated as 

\begin{equation}
    \hat y = \sum_{i=1}^{N} \alpha_i \kappa(\boldsymbol{x},\boldsymbol{x}_i),
\end{equation}
where $\alpha$ is the dual parameter, $\kappa$ is the kernel function and $N$ denotes number of sample points. 
The dual parameters for both models are calculated as

\begin{alignat*}{3}
    \alpha &= (\boldsymbol{\mathrm{K}} + {\sigma_n}^2\boldsymbol{\mathrm{I}} )^{-1} \boldsymbol{y} &\vspace*{1cm} \rightarrow \text{GP Regressor}\\
    \alpha &= (\boldsymbol{\mathrm{K}} + \lambda \boldsymbol{\mathrm{I}})^{-1} \boldsymbol{y} &\vspace*{1cm} \rightarrow \text{Kernel Ridge}
\end{alignat*}
where $\boldsymbol{\mathrm{K}}$ is the kernel matrix, $\lambda$ is the regularization term, $\boldsymbol{\mathrm{I}}$ is 
the identity matrix, and ${\sigma_n}^2$ is the noise variance. As long as the same kernel function with identical kernel 
parameters is used and the condition $\lambda={\sigma_n}^2$ is met, the mean predictions of the GP Regressor closely 
follow the predictions of Kernel Ridge. 
To verify this, two cases are considered. 
In the first case, a GP Regressor using the Radial Basis Function (RBF) kernel with the length scale 
of 1 and ${\sigma_n}^2=\SI{1e-10}{}$, along with a Kernel Ridge using the same kernel with $\lambda=\SI{1e-10}{}$, are trained.
In this case, both models demonstrate strong alignment between observed data and their respective predictions. 
Additionally, both predictions closely match each other, as depicted in Figure~\ref{fig:figure1}. In the second case,
illustrated in Figure~\ref{fig:figure2},  
a GP Regressor using the DotProduct kernel with ${\sigma_n}^2=\SI{0}{}$ (an equivalent representation of a linear kernel) is 
compared to a Kernel Ridge employing the linear kernel. 
Here, although there is a noticeable misalignment between observed data and predictions, 
the predictions of both models closely match each other, resembling the behavior observed in the first case. Moreover, 
in both scenarios, point-wise errors between predictions of GP Regressor and Kernel Ridge confirm these observations. 
Note that the kernel used in the first case is the kernel of the baseline GP Regressor, while the kernel used in the 
second case is the kernel of the baseline Kernel Ridge (see Figure~\ref{fig:baseline_comparison}). Therefore, we can see 
similar behavior on effects on the predictions.

\end{appendices}


\bibliography{ml_dl_for_rsc}

\end{document}